\documentclass[aps, prd, amsmath, floats, floatfix, twocolumn,
superscriptaddress, nofootinbib, showpacs]{revtex4}
\usepackage{graphicx}
\usepackage{epsfig}
\usepackage{color}
\usepackage{soul}
\usepackage{url}
\usepackage{bm}         % bold math symbols
\usepackage{times}

\newcommand{\beq}{\begin{equation}}
\newcommand{\eeq}{\end{equation}}
\newcommand{\beqn}{\begin{eqnarray}}
\newcommand{\eeqn}{\end{eqnarray}}

\newcommand{\lo}{\mathrel{\raise.3ex\hbox{$<$}\mkern-14mu
    \lower0.6ex\hbox{$\sim$}}}
\newcommand{\go}{\mathrel{\raise.3ex\hbox{$>$}\mkern-14mu
    \lower0.6ex\hbox{$\sim$}}}

\usepackage{color}

\newcommand{\Caltech}{\affiliation{Theoretical Astrophysics 350-17,
    California Institute of Technology, Pasadena, California 91125, USA}}
\newcommand{\Cornell}{\affiliation{Center for Radiophysics and Space
    Research, Cornell University, Ithaca, New York, 14853, USA}}
\newcommand{\WSU}{\affiliation{Department of Physics \& Astronomy,
	Washington State University, Pullman, Washington 99164, USA}}
\newcommand{\CITA}{\affiliation{Canadian Institute for Theoretical 
    Astrophysics, University of Toronto, Toronto, Ontario M5S 3H8, Canada}}

\newcommand{\CIFAR}{\affiliation{Canadian Institute for Advanced Research, 180 Dundas St.~West, Toronto, ON M5G 1Z8, Canada}} %

\usepackage{graphicx}% Include figure files
\usepackage{dcolumn}% Align table columns on decimal point
\usepackage{bm}% bold math
\usepackage{epsf}

\begin{document}

\title{Neutron star-black hole mergers with a nuclear equation of state and neutrino cooling:\\
 Dependence in the binary parameters}

\author{Francois Foucart} \CITA%
\author{M. Brett Deaton} \WSU %
\author{Matthew D. Duez} \WSU %
\author{Evan O'Connor} \CITA%
\author{Christian D. Ott} \Caltech 
\author{Roland Haas} \Caltech
\author{Lawrence E. Kidder} \Cornell %
\author{Harald P. Pfeiffer} \CITA\CIFAR
\author{Mark A. Scheel} \Caltech %
\author{Bela Szilagyi} \Caltech %

\begin{abstract}
We present a first exploration of the results of neutron star-black hole mergers using black hole masses in the most likely range
of $7M_\odot\textrm{--}10M_\odot$, a neutrino leakage scheme, and a modeling of the neutron star material through a finite-temperature nuclear-theory based
equation of state. In the range of black hole spins in 
which the neutron
star is tidally disrupted ($\chi_{\rm BH}\gtrsim 0.7$), we show that the merger consistently produces large amounts of cool ($T\lesssim 1\,{\rm MeV}$), 
unbound, neutron-rich material ($M_{\rm ej}\sim 0.05M_\odot\textrm{--}0.20M_\odot$).
A comparable amount of bound matter is initially divided between a hot disk ($T_{\rm max}\sim 15\,{\rm MeV}$) 
with typical neutrino luminosity $L_\nu\sim 10^{53}\,{\rm erg/s}$, and a cooler tidal tail. 
After a short period of rapid protonization
of the disk lasting $\sim 10\,{\rm ms}$, the accretion disk cools down under the combined effects of 
the fall-back of cool material from the tail, continued accretion of the hottest material onto the black hole,
and neutrino emission.
As the temperature decreases, the disk progressively becomes more neutron-rich, with dimmer neutrino emission. 
This cooling process should
stop once the viscous heating in the disk (not included in our simulations) balances the cooling. 
These mergers of neutron star-black hole binaries with black hole masses $M_{\rm BH}\sim 7M_\odot\textrm{--}10M_\odot$ and black hole
spins high enough for the neutron star to disrupt provide promising candidates for the production of short gamma-ray bursts,
of bright infrared post-merger signals due to the radioactive decay of unbound material, and of large amounts of r-process
nuclei.
\end{abstract}


\maketitle

\section{Introduction}
\label{sec:intro}

The coalescence and merger of black holes and neutron stars in binary systems is one of the main sources of gravitational waves
that is expected to be detected by the next generation of ground-based detectors (Advanced LIGO~\cite{Harry2010} and VIRGO~\cite{aVIRGO}, KAGRA~\cite{Somiya:2011np}). 
If at least one of the members of the binary is a neutron star, bright post-merger electromagnetic signals could also be observable. For
example, the formation of hot accretion disks around remnant black holes provides a promising setup for the generation of short gamma-ray bursts
(SGRB), while the radioactive decay of unbound neutron-rich material could power an infrared transient days after the merger 
(`kilonova')~\cite{metzger:11}. Numerical simulations of these mergers in a general relativistic framework are required
both to model the gravitational wave signal around the time of merger, and to determine which binaries can produce detectable electromagnetic
signals and how the properties of these signals are related to the physical parameters of the source. 

These two objectives demand very different types of simulations. To model the gravitational wave signal, 
long and very accurate simulations of the last tens of orbits before merger are required. 
But during this phase, the important physical effects can be recovered using simple models
for the neutron star matter (e.g.\ Gamma-law, piecewise polytrope)~\cite{Read:2008iy,Lackey2011}.
On the other hand, to assess whether a given binary can power detectable electromagnetic signals
and to predict nucleosynthesis yields, 
shorter inspiral simulations are acceptable but a detailed
description of a wider array of physical effects is required: magnetic fields, neutrino radiation, nuclear reactions and the composition 
and temperature dependence of the properties of neutron-rich, high-density material all play an important role in the post-merger evolution
of the system. Any ejected material also has to be tracked far from the merger site, thus requiring accurate evolutions of the fluid in a much larger
region than during inspirals. The simulations presented here focus on the second issue in the context of neutron star-black hole (NSBH) mergers.

General relativistic simulations have only recently begun to include the effects
of neutrino radiation on the post-merger evolution of the remnant of  binary neutron star (BNS)~\cite{Sekiguchi:2011zd,Kiuchi:2012mk,2014arXiv1403.3680N} and
NSBH~\cite{Deaton2013} mergers through the use of `leakage' schemes providing a simple prescription for the cooling
of the fluid through neutrino emission. These schemes are largely based on methods already used in the simulation of post-merger remnants by codes
with approximate treatments of gravity~\cite{Ruffert1996,1997A&A...319..122R,Rosswog:2003rv,Setiawan2006}. More advanced (and computationally expensive)
methods based on a moment expansion of the radiation fields~\cite{1981MNRAS.194..439T} have also been developed for general relativistic 
simulations of binary mergers~\cite{2011PThPh.125.1255S}. An energy-integrated version of the moment formalism was 
recently used for the first time to study BNS mergers~\cite{2014arXiv1402.7317W}.
Both general relativistic and non-relativistic simulations have shown that neutrino cooling can play an important role in the evolution of the
disk. Emission and absorption of neutrinos can also affect the composition of the ejecta, and the mass of heavy elements produced as a result
of r-process nucleosynthesis in the unbound material~\cite{2008ApJ...679L.117SR,Roberts2011,2014arXiv1402.7317W}. 
Finally, energy deposition by neutrino-antineutrino annihilation in the baryon-poor region above the disk and near the poles of the black hole
could play a role (either positive or negative) in the production of short gamma-ray bursts~\cite{2007NJPh....9...17L}.

Magnetic fields are also expected to play a critical role in the
evolution of NSBH remnants.  Simulations show magnetic effects to be
unimportant before merger for realistic field strengths~\cite{Chawla:2010sw,
2012PhRvD..85f4029E}.  However, if the merger produces an accretion disk,
the disk will be subject to the magneto-rotational instability (MRI)~\cite{BalbusHawley1991}, which
will induce turbulence, leading to angular momentum transport and energy
dissipation that drives the subsequent accretion.  Most NSBH simulations
with magnetic fields fail to resolve MRI growth, but it is seen if a
sufficiently strong poloidal seed field is inserted~\cite{Etienne:2012te}.

Finally, the equation of state of the fluid used to model the neutron star plays an important role before, during and after a NSBH
merger. Before merger, it determines the response of the neutron star to the tidal field of the black hole, which,
for low mass black holes, could cause measurable differences in the gravitational wave signal~\cite{Lackey2011,Lackey:2013}. During merger,
it determines whether the neutron star is disrupted by the black hole (mostly by setting the radius of the neutron star), 
allowing the formation of an accretion disk and the ejection of
unbound material, or whether it just falls directly into the black hole. Different equations of state can also lead to different qualitative
features for the evolution of the tidally disrupted material. After merger, knowing the temperature and composition dependence of the equation of
state is necessary to properly evolve the forming accretion disk. 
%The equation of state of nuclear-density matter is not currently known,
%and is in fact one of the parameters that we can hope to constrain by observing binary mergers. 
Until recently, most general relativistic 
simulations used Gamma-law or piecewise-polytropic equations of state, which can only provide us with accurate results up to the 
disruption of the neutron
star. For post-merger evolution, composition- and temperature-dependent
nuclear-theory based equations of state are required both to properly model the properties of the fluid
and to be able to compute its composition and its interaction with neutrinos. 
Only two general relativistic simulations of NSBH
mergers using such equations of state have been presented so far. First, a single low-mass case without neutrino radiation~\cite{Duez:2009yy} showed
only moderate differences with an otherwise identical simulation using a simpler Gamma-law equation of state. More
recently, a first simulation with our neutrino
leakage scheme for a relatively low mass, highly spinning black hole~\cite{Deaton2013} indicated that the effects of neutrino cooling were 
more important than the details of the equation of state.
Simulations using zero-temperature, nuclear-theory based equations of state were also reported as part of a couple of studies of the radioactive
emission coming from unbound neutron-rich material~\cite{2013ApJ...778L..16H,2014ApJ...780...31T}, 
but only a few general properties of the ejecta have been provided at this point. 
Accordingly, our best estimates for the dependence of the merger and post-merger evolution of NSBH binaries
on the equation of state of nuclear matter come from a set of studies using simpler models for the nuclear 
matter~\cite{Duez:2009yy,Kyutoku:2010zd,Kyutoku:2011vz}, which need to be complemented by simulations using
temperature-dependent, nuclear-theory based equations of state.

In this paper, we study with the SpEC code ~\cite{SpECwebsite}
the merger and post-merger evolution of NSBH binaries for black holes of mass
$M_{\rm BH}=7M_\odot\textrm{--}10M_\odot$ (around the current estimates of the peak of the mass distribution of stellar mass black 
holes~\cite{Ozel:2010,Kreidberg:2012}). We use a nuclear-theory based equation of state with temperature and composition dependence
of the fluid properties, and a leakage scheme to approximate the effects of neutrino cooling. 
Given the computational cost of the numerical simulations, we limit ourselves to a single 
equation of state (LS220 equation of state from Lattimer \& Swesty~\cite{Lattimer:1991nc}) and only consider relatively low mass neutron 
stars ($M_{\rm NS}=1.2M_\odot\textrm{--}1.4M_\odot$). We also use black holes with spins high enough for the neutron star to be tidally
disrupted before plunging into the black hole (dimensionless spins $\chi_{\rm BH}=0.7\textrm{--}0.9$, as estimated from simulations using
simpler equations of state~\cite{Foucart2012}). Indeed, for lower spin black holes, the post-merger evolution is trivial
and we do not expect the production of appreciable nucleosynthesis output or detectable electromagnetic signals. 
For the same black hole masses $M_{\rm BH}=7M_\odot-10M_\odot$ but higher black hole spins, on the other hand,
the neutron star is tidally disrupted during the merger.
We show that large amounts of neutron-rich, low entropy material are ejected, which will undergo robust r-process nucleosynthesis.
Bound material which escapes rapid accretion onto the black hole forms an accretion disk, albeit generally of slightly lower mass 
than for binaries of more symmetric mass ratio and comparable black hole spins (we find $M_{\rm disk}\sim 0.05M_\odot\textrm{--}0.15M_\odot$).
The disk is initially hot ($T_{\rm max}\sim 15\,{\rm MeV}$), with a high neutrino luminosity ($L_\nu \sim 10^{53}{\rm erg/s}$). 
But about $10\,{\rm ms}$ after merger, the combined effect of the accretion of hot material by the black hole,
the fall-back of cold material from the tidal tail, the emission of neutrinos, and the expansion of the disk 
causes a rapid decline of both the temperature and the luminosity. 
In a realistic disk, in the inner regions where neutrino cooling is efficient, this decline would presumably stop when the neutrino cooling 
and the viscous heating due to MRI turbulence roughly balance each other.
However, our simulations at this point do not include magnetic effects, and would
not resolve the MRI even if they did. And we do not include any parametrized viscosity either. Accordingly, here the disk continues to cool until we end the simulation, 
about $40\,{\rm ms}$ after merger. 
With both a massive, hot accretion disk and a significant amount of ejected unbound material, NSBH mergers in this part
of the parameter space would thus be prime candidates for the emission of both prompt (e.g. short gamma-ray bursts) and delayed (e.g. `kilonovae')
electromagnetic signals.

We find that the impact of using the LS220 equation of state appears, for the lower mass black holes, fairly weak before merger. 
Most of the difference in post-merger evolution compared to previous simulations using neutron stars of 
similar radii but simpler equations of state 
can be attributed to the effect of neutrino cooling. For higher mass black holes, on the other hand, the disruption occurs
very close to the black hole horizon, and the detail of the tidal response of the neutron star can cause more significant differences.
The disruption occurs later, faster, and creates a much narrower tidal tail for the LS220 equation of state than for
the commonly used Gamma-law equation of state with $\Gamma=2$. These differences also cause the disruption of the neutron star
to be extremely hard to resolve numerically, and require the use of a much finer grid close to the black hole horizon than what is typically
used in general relativistic simulations of NSBH mergers (see Sec.~\ref{sec:numsetup}).

We will begin with a discussion of the numerical methods used, the chosen initial configurations and an estimate of the errors 
in the simulations (Sec.~\ref{sec:numsetup}). We then discuss the general properties of the disruption and the evolution 
of the accretion disk (Sec.~\ref{sec:dynamics}), before a more detailed presentation of the neutrino emission from the disk, 
and of the evolution of its composition (Sec.~\ref{sec:neutrino}), and a summary of the expected evolution
of the disk over timescales longer than our simulations (Sec.~\ref{sec:ltdisk}). 
Finally, Sec.~\ref{sec:waves} briefly discusses the gravitational wave signal.

%\subsection{Unit conventions}
%
Unless units are explicitly given, we use the convention $G=c=1$, where $G$ is the gravitational constant and $c$ the
speed of light. 

\section{Numerical Setup}
\label{sec:numsetup}

\subsection{Equation of state}
\label{sec:eos}

\begin{figure}
\flushleft
\includegraphics[width=0.85\columnwidth]{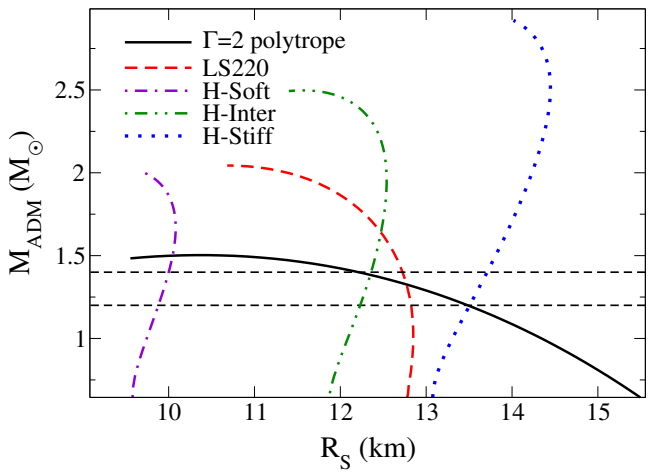}
\caption{Mass-radius relationship for the LS220 equation of state, the 3 equations of state
presented in Hebeler et al.~\cite{2013ApJ...773...11H}, and a $\Gamma=2$ polytrope of similar radius for a $1.4M_\odot$
star. We consider the ADM mass of isolated neutron stars, and their areal radius. 
Dashed lines indicate the masses used in this work. The H-Soft and H-Stiff equations of state 
bracket the ensemble of mass-radius
relationships obtained for the family of equations of state presented in~\cite{2013ApJ...773...11H}. 
These equations of state match both 
neutron star mass constraints and nuclear physics constraints from chiral effective field theory. H-Inter 
is a representative intermediate case.}
\label{fig:MR}
\end{figure}

\begin{figure}
\flushleft
\includegraphics[width=0.85\columnwidth]{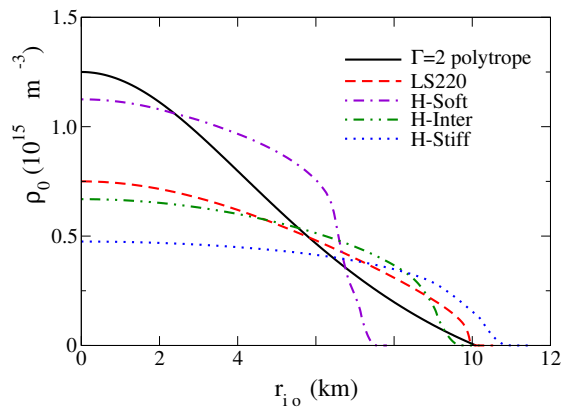}
\caption{Density profile as a function of radius (in isotropic coordinates) for the LS220 equation of state, the 3 equations of state
presented in Hebeler et al.~\cite{2013ApJ...773...11H}, and a $\Gamma=2$ polytrope of similar radius for a $1.4M_\odot$
star. Note that in isotropic coordinates, the radius of the surface is not equal to the circumferential radius.}
\label{fig:Rho}
\end{figure}

As in our first study of NSBH mergers including neutrino cooling~\cite{Deaton2013}, we model
the nuclear matter using the Lattimer \& Swesty 
equation of state~\cite{Lattimer:1991nc}
with nuclear incompressibility parameter $K_0=220\,{\rm MeV}$ and symmetry
energy $S_\nu=29.3\,{\rm MeV}$ (hereafter LS220), 
using the table available on http://www.stellarcollapse.org
and described in O'Connor \& Ott 2010~\cite{OConnor2010}. 
This equation of state lies within the allowed range of neutron star radii,
as determined by Hebeler et al.~\cite{2013ApJ...773...11H}
from nuclear theory constraints and the existence of 
neutron stars of mass $\sim 2M_\odot$~\cite{Demorest:2010bx,2013Sci...340..448A}. 
Although it is not fully consistent with the most recent
constraints from nuclear experiments (in particular, measurements of the Giant
Dipole Resonance~\cite{2013ApJ...771...51L,2013ApJ...774...17S}), the
%properties of a neutron star using the LS220 equation of state which are 
%most relevant to our merger simulations appear close to those of
%more realistic equations of state (see Figs.~\ref{fig:MR}-\ref{fig:Rho}).
%Using the LS220 equation of state is, at the very least, a significant
%step forward from ideal gas equations of state, or even from the
%nuclear theory based equation of state first used in NSBH 
%simulations~\cite{Shen:1998gq,Duez:2009yy}.
LS220 equation of state produces neutron stars over a range of
masses with structural properties within the limits set by chiral
effective field theory (see Figs.~\ref{fig:MR}-\ref{fig:Rho}).
Using the LS220 equation of state is, at the very least, a
significant step forward from Gamma-law equations of state, or
even from the temperature-dependent nuclear-theory based 
equation of state first used in
NSBH simulations~\cite{Shen:1998gq,Duez:2009yy}, which do not meet
these constraints.
Tabulated equations of state which do satisfy all known constraints
have recently been developed (see e.g.~\cite{2013ApJ...774...17S})
and are available in tabulated form on stellarcollapse.org. Those
equations of state will be the subject of upcoming numerical
studies.

\begin{table}
\caption{
  Properties of the neutron stars before tidal effects become strong. 
$M^b_{\rm NS}$ is the baryonic mass, $M_{\rm NS}$ the ADM mass of the star, if
isolated, $R_{\rm NS}$ the circumferential radius, and
$C_{\rm NS}\equiv M_{\rm NS}/R_{\rm NS}$ the compactness.
}
\label{tab:NS}
\begin{tabular}{|c|c|c|c|}
\hline
$M_{\rm NS}$ & $R_{\rm NS}$ & $C_{\rm NS}$ & $M^b_{\rm NS}$ \\
\hline
$1.20M_\odot$ & ${12.8\,\rm km}$ & $0.139$ & $1.31M_\odot$ \\
$1.40M_\odot$ & ${12.7\,\rm km}$ & $0.163$ & $1.55M_\odot$ \\
\hline
\end{tabular}
\end{table}

Table~\ref{tab:NS} summarizes the properties of the two neutron stars that we use
in this work (with masses $M_{\rm NS}=1.2M_\odot$ and $M_{\rm NS}=1.4M_\odot$). With
radii of $\sim 12.7\,{\rm km}$, they lie within the range of
sizes allowed by chiral effective field theory constraints
(see Fig.~\ref{fig:MR})~\cite{2013ApJ...773...11H}.
The most recent measurements of neutron star radii in X-ray binaries~\cite{2013ApJ...765L...5S}
and the millisecond pulsar PSR J0437--4715~\cite{2013ApJ...762...96B} are also consistent
with $R_{\rm NS}\sim 12.7\,{\rm km}$ (but see~\cite{Guillot:2013wu} for predictions
of more compact neutron stars).

%\begin{figure}
%\flushleft
%\includegraphics[width=0.85\columnwidth]{hvsrho}
%\caption{Enthalpy per unit mass as a function of the baryon density for the LS220 equation of state, the intermediate equation of state
%presented in Hebeler et al.~\cite{2013ApJ...773...11H}, and a $\Gamma=2$ polytrope of similar radius for a $1.4M_\odot$
%star.}
%\label{fig:hvsrho}
%\end{figure}

%Finally, in Fig.~\ref{hvsrho}, we plot the specific enthalpy (per unit mass) of the fluid as a function of the baryon density for 
%the LS220 equation of state, the closest equation of state from~\cite{2013ApJ...773...11H}, and a $\Gamma=2$ polytrope of similar
%radius. This quantity is important as it determines the behavior of the fluid during decompression (see Sec.~\ref{sec:ejecta}).
%Both nuclear-theory based equations of state are characterized by a stiff core (when neutron degeneracy pressure becomes important)
%and a softer atmosphere, while the polytrope as similar properties at all densities.

\subsection{Initial configurations}
\label{sec:ID}

The initial conditions for our simulations are chosen so that the merger leads 
to the disruption of the neutron star, and thus potentially to the 
formation of an accretion disk and the ejection of unbound material. 
We additionally require the black hole mass
to be within the range currently favored by observations of galactic black 
holes~\cite{Ozel:2010,Kreidberg:2012} and by population synthesis 
models~\cite{belczynski:08,2010ApJ...715L.138B}, $M_{\rm BH}\sim 7M_\odot-10M_\odot$.
In this mass range, the neutron star will only be disrupted for rapidly spinning
black holes and large neutron stars. An approximate threshold for the disruption of the 
neutron star to occur is indeed~\cite{Foucart2012}
\beq
C_{\rm NS} \lesssim \left(2+2.14 q^{2/3} \frac{R_{\rm ISCO}}{6M_{\rm BH}}\right)^{-1}\,\,,
\eeq
where $C_{\rm NS}=M_{\rm NS}/R_{\rm NS}$ is the compactness of the star, $q=M_{\rm BH}/M_{\rm NS}$ 
is the mass ratio, and $R_{\rm ISCO}$ is the radius of the innermost stable circular orbit
for an isolated black hole of the same mass and spin.
For a canonical neutron star of mass $1.4M_\odot$ described by the LS220 equation of state,
this requires a dimensionless spin $\chi_{\rm BH}\gtrsim 0.75$ (resp.\ $\chi_{\rm BH}\gtrsim 0.55$) 
for a black hole of mass $10M_\odot$ (resp.\ $7M_\odot$). For the LS220 equation of state,
the critical spin separating disrupting from non-disrupting neutron stars increases with
the neutron star mass. Accordingly, we will consider black hole spins in the range
$\chi_{\rm BH}=0.7\textrm{--}0.9$ and neutron star masses in the range $M_{\rm NS}=1.2M_\odot\textrm{--}1.4M_\odot$.

The simulations presented in this paper are fairly costly: at our standard resolution,
a single run might require $100,000\textrm{--}150,000$ CPU-hours (more than 3 months using 48 processors).
As we are considering a 3-dimensional parameter space (neutron star and black hole mass,
and black hole spin magnitude), we can only afford a very coarse coverage of each parameter.
In this first parameter space study using the LS220 equation of state, we consider 9 simulations
covering 2 black hole masses ($7M_\odot$, $10M_\odot$), 2 neutron star masses ($1.2M_\odot,1.4M_\odot$),
and 3 spins ($\chi_{\rm BH}=0.7,0.8,0.9$, the lower spin being only used for $M_{\rm NS}=1.4M_\odot$, $M_{\rm BH}=7M_\odot$).
The initial parameters for each configuration are summarized in Table~\ref{tab:ID}, while
the properties of the neutron star for each choice of $M_{\rm NS}$ are given in Table~\ref{tab:NS}.
For reference, our first simulation using the LS220 equation of state and a leakage scheme for the
neutrino radiation used a $M_{\rm NS}=1.4M_\odot$ neutron star, with a less massive ($M_{\rm BH}=5.6M_\odot$),
rapidly rotating ($\chi_{\rm BH}=0.9$) black hole~\cite{Deaton2013}. In the following sections, we will refer to the various
simulations by their names listed in Table~\ref{tab:ID}, in which the first two numbers refer to the mass
of the neutron star and the mass of the black hole, and the third number to the spin of the black hole, e.g.\ M12-10-S9
correspond to a binary with $M_{\rm NS}=1.2M_\odot$, $M_{\rm BH}=10M_\odot$ and $\chi_{\rm BH}=0.9$.

\begin{table}
\caption{
Initial configurations studied. All cases use the LS220 equation of state.
$M_{\rm BH}$ and $\chi_{\rm BH}$ are the mass and dimensionless
spin of the black hole, $M_{\rm NS}$ the mass of the neutron star,
$M\Omega_{\rm orbit}^0$ the initial orbital frequency multiplied
by the total mass $M=M_{\rm BH}+M_{\rm NS}$, and $e$ is the
initial eccentricity.
}
\label{tab:ID}
\begin{tabular}{|c||c|c|c|c|c|}
\hline
Name & $M_{\rm BH}$ & $\chi_{\rm BH}$ & $M_{\rm NS}$ & $M\Omega_{\rm orbit}^0$ & $e$ \\
\hline
M12-7-S8 & $7M_\odot$ & 0.8 & $1.2M_\odot$ & 0.0364 & 0.027 \\ 
M12-7-S9 & $7M_\odot$ & 0.9 & $1.2M_\odot$ & 0.0364 & 0.026 \\
M12-10-S8 & $10M_\odot$ & 0.8 & $1.2M_\odot$ & 0.0396 & 0.031 \\
M12-10-S9 & $10M_\odot$ & 0.9 & $1.2M_\odot$ & 0.0396 & 0.033\\
M14-7-S7 & $7M_\odot$ & 0.7 & $1.4M_\odot$ & 0.0438 & 0.039 \\
M14-7-S8 & $7M_\odot$ & 0.8 & $1.4M_\odot$ & 0.0437 & 0.037 \\
M14-7-S9 & $7M_\odot$ & 0.9 & $1.4M_\odot$ & 0.0437 & 0.037 \\
M14-10-S8 & $10M_\odot$ & 0.8 & $1.4M_\odot$ & 0.0441 & 0.042 \\
M14-10-S9 & $10M_\odot$ & 0.9 & $1.4M_\odot$ & 0.0440 & 0.043 \\
\hline
\end{tabular}
\end{table}

We obtain constraint satisfying initial data using the spectral elliptic solver Spells~\cite{Pfeiffer2003,FoucartEtAl:2008},
at a separation chosen to provide $5\textrm{--}8$ orbits before merger. The initial data use the quasi-circular 
approximation~\cite{Taniguchi:2005fr,FoucartEtAl:2008}, which
causes the binary to be on slightly eccentric orbits ($e\sim 0.03\textrm{--}0.04$, see Table~\ref{tab:ID}).

\subsection{Summary of the neutrino leakage scheme}
\label{sec:nuleak}

The neutrino leakage scheme used in this work is a first attempt at including the effects of neutrino
radiation on the evolution of the remnant of NSBH mergers and at providing estimates of the
properties of the emitted neutrinos (luminosity, species, average energy). 
A leakage scheme is a local prescription for the number and energy of neutrinos
emitted from a given point in the disk, based on the local properties of the fluid and on an estimate
of the optical depth. A detailed description of our leakage implementation can be found in 
Deaton et al. (2013)~\cite{Deaton2013}. It is essentially a generalization to 3 dimensions of the 
leakage code implemented in GR1D (O'Connor \& Ott 2010~\cite{OConnor2010}), which is itself inspired by the
work of Ruffert et al. (1996)~\cite{Ruffert1996} and Rosswog \& Liebend{\"o}rfer (2003)~\cite{Rosswog:2003rv}.
Our leakage scheme aims at modeling the effects of neutrino cooling, while we do not
consider the effects of neutrino heating.

As a brief summary, the main components of the leakage code are:
\begin{itemize}
\item A local prescription for the free emission of neutrinos as a function of the fluid properties (i.e.\ 
emission in the optically thin regime). We include $\beta$-capture processes, $e^+$-$e^-$ pair annihilation,
plasmon decay~\cite{Ruffert1996} and nucleon-nucleon Bremsstrahlung~\cite{Burrows2006}. We compute the
luminosity and number emission for $\nu_e$, $\bar{\nu}_e$ and $\nu_x$, where $\nu_x$ stands for all
other neutrino species ($\nu_{\mu,\tau},\bar{\nu}_{\mu,\tau}$), which are assumed to all have the same
interactions with the fluid.
\item A local prescription to compute the opacity to neutrinos as a function of the fluid properties and of 
the chemical potentials. We take into account scattering on nucleons and heavy nuclei and absorptions
on nucleons~\cite{OConnor2010}.
\item A prescription for the computation of the optical depth between the current point and the boundaries
of the domain. We estimate this by integrating the opacities along the coordinate axes, and along the two most
promising diagonal directions (i.e. those corresponding to the coordinate axes along which the optical depth
is minimum). The minimum optical depth along the selected path is taken as the optical depth at the current
point (see Sec.~\ref{sec:errors} for an estimate of the influence of this choice).
\item A local prescription for the rate at which neutrinos can escape through diffusion (i.e.\ in the optically
thick regime), for which we follow Rosswog \& Liebend{\"o}rfer (2003)~\cite{Rosswog:2003rv}.
\item A choice of an effective emission rate interpolating between the optically thin and optically thick limit.
If $Q_{\rm free}$ is the emission rate in the free emission regime and $Q_{\rm diff}$ the emission rate in the
diffusive regime, we use 
\beq
Q_{\rm eff} = \frac{Q_{\rm diff}Q_{\rm free}}{Q_{\rm free}+Q_{\rm diff}}
\eeq  
(for both the energy and number emission rates).
\end{itemize}

Compared to Deaton et al. 2013~\cite{Deaton2013}, we have implemented two modifications to the leakage scheme.
First, we compute the neutrino chemical potentials used in the determination of the opacities
\beq
\mu_\nu = \mu_\nu^{\rm eq} (1 - e^{-\langle \tau_\nu\rangle})\,\,,
\eeq
where $\mu_\nu^{\rm eq}$ is the $\beta$-equilibrium value of the potential and $\langle\tau_\nu\rangle$ the energy-averaged
optical depth, by iteratively solving for ($\mu_\nu$,$\tau_\nu$). This is required because $\tau_\nu$ is itself a function
of the opacities. In~\cite{Deaton2013}, we instead used an analytical approximation for $\langle\tau_\nu\rangle$ as a function
of the local properties of the fluid when computing $\mu_\nu$.  
Second, instead of treating the black hole as an optically thick region, we let neutrinos freely escape through
the excision boundary. In~\cite{Deaton2013}, the black hole was treated as an optically thick region
to avoid spuriously including the emission from hot points at the inner edge of the disk in the
total neutrino luminosity. In this paper, when computing the optical depth along a direction 
crossing the excision boundary, we terminate the integration at that boundary. But points for which the 
direction of smallest optical depth is towards the black hole are excluded when computing the total 
neutrino luminosity of the disk. The cooling and composition evolution of the fluid are thus properly
computed at the inner edge of the disk, but without affecting estimates of the neutrino luminosity.
These changes neither appear to significantly modify the results of the simulations at late times, nor the 
properties or evolution of the post-merger accretion disk.

\subsection{Error estimates}
\label{sec:errors}

To estimate the errors in the various observables discussed in this paper for simulations with 
$M_{\rm BH}=7M_\odot$, we perform a convergence test on
simulation M14-7-S9, as well as a series of tests checking that the treatment of the low-density regions,
the boundary condition on the region excised from the numerical domain inside of the apparent horizon 
of the black hole, and the method used to compute the neutrino optical depth do not affect 
the results within the expected numerical
errors. For the simulations with $M_{\rm BH}=10M_\odot$, which are significantly harder to resolve, we also
simulate the late disruption phase using a fixed mesh refinement and a much finer grid and find that 
the standard grid choices used for the lower mass ratio cases are no longer appropriate in this regime. 
Issues specific to the high mass ratio cases are discussed at the end of this section.

For the convergence test with $M_{\rm BH}=7M_\odot$, we perform additional simulations at a lower and a higher resolutions. 
The SpEC code~\cite{SpECwebsite} uses two separate numerical grids~\cite{Duez:2008rb}: 
a finite difference grid on which we evolve the general relativistic fluid 
equations, and a pseudospectral grid on which we evolve Einstein's equations. The finite difference grid is regularly regenerated so that
it covers the region in which matter is present, but ignores the rest of the volume 
covered by the pseudospectral grid. The coupling between the two sets of equations
then requires interpolation from one grid to the other, which we perform at the end of each time step.
During the plunge and merger,
the three resolutions have $N_{\rm FD}=(120^3,140^3,160^3)$ points on the finite difference grid, which only covers
the region in which matter is present. This leads to variation of the physical grid spacing over time, from $\Delta x\sim 200\,{\rm m}$
at the beginning of the plunge to $\Delta x\sim 2\,{\rm km}-3\,{\rm km}$ while we follow the ejected material as
it escapes the grid, to finally $\Delta x\sim 1\,{\rm km}$ at the end of the simulation. 
The resolution on the spectral grid on which we evolve Einstein's equations is chosen adaptively,
so that the relative truncation error of the spectral expansion of the metric and of its spatial derivatives is kept
below $\epsilon_{\rm sp}=(1.9\times 10^{-4},1\times 10^{-4}, 5.9\times 10^{-5})$ at all times (as measured from
the amplitude of the coefficients of the spectral expansion within each subdomain of the numerical grid). 
During the short inspiral, the finite difference grid covers a 
much smaller region (a box of size $2R_{\rm NS}\sim 26\,{\rm km}$) and has slightly less points ($N_{\rm FD}=(100^3,120^3,140^3)$). The truncation error on the spectral
grid is kept at about the same level, except close to the black hole and neutron star where we require the truncation
error to be a factor of 10 smaller. At these resolutions, we find that for many quantities the convergence is faster than second
order, which is the expected convergence rate in the high resolution limit when finite difference errors dominate the error
budget. This is presumably due to contributions to the error of terms with a higher convergence rate, such as grid-to-grid interpolation
(third order), the evolution of Einstein's equations (exponentially convergent everywhere but in 
regions in which a discontinuity is present), or the time-stepping algorithm (third order). 
When we observe a convergence rate faster than second-order, we assume second order convergence at the two highest resolutions
to get an upper bound on the error. In previous simulations with the SpEC code, we have generally found this error estimate 
(which is about 4 times the difference between the results of the high and medium resolutions) to be much higher
than the actual error in the results. Using this estimate, we find the following upper bounds for the errors in our medium resolution runs:
\begin{itemize}
\item $\sim 25\%$ relative error in the final disk mass %(a difference slightly below $0.01M_\odot$ between the medium and high resolutions);
\item $\sim 60\%$ relative error in the mass and kinetic energy of the ejected material;
%(with the same absolute difference between medium and high resolutions as for the 
%final disk mass), as well as its kinetic energy.  
\item $1\%$ relative error in the mass and spin of the black hole at a given time; 
%(and also $\sim 1\%$ changes in these properties at late times
%due to continued accretion onto the black hole from the disk).
\item $\sim 50\%$ relative error in the neutrino luminosity obtained from the leakage scheme -- smaller than the error due
to the use of the leakage algorithm itself, which might be a factor of a few (see e.g.~\cite{Rosswog:2003rv}).
\item $\sim 10\,{\rm km/s}$ of error in the recoil velocity due to gravitational wave emission ($v_{\rm kick}^{\rm GW}$
in Table~\ref{tab:remnant}), due mostly to the extrapolation of the waveform to infinity.
%\item $\sim 15\%$ in the relative error in the average energy of neutrinos at a given time.
\end{itemize}

Beyond the finite resolution of the numerical grids, our choice of numerical methods cause additional errors which are
important to quantify. First, we usually stop evolving the metric variables $\sim 15\,{\rm ms}$ after merger. By that point,
the mass of the post-merger disk is at most $\sim 1\%$ of the mass of the black hole. Previous tests on configurations
with a relatively higher disk mass ($M){\rm disk}\sim 0.03 M_{\rm BH}$) have however shown that, at the level of accuracy
currently reached by our simulations, this has no observable effect on the results~\cite{Foucart:2010eq}. Similary,
$\sim 1\%-2\%$ of the mass of the system leaves the grid as unbound or marginally bound material. By performing simulations
in which the matter evolution was followed up to different radii, we have confirmed that the error thus induced (both in
the evolution of the post-merger remnant and the evolution of Einstein's equation) is not measurable at our current accuracy,
up to the point at which fall-back of the neglected bound material would affect the evolution of the disk. But this will only
occur after multiple orbital timescale of the disk, by which point the fact that we neglect the effective viscosity induced
by the growth of the magnetorotational instability is expected to be a much larger source of error.

In the leakage scheme, one potential source of error is the choice of specific directions over which the optical depth is computed
(in our case, the coordinate axes and a subset of the diagonals). To assess the effect of that choice, we tried two alternative
methods. In the first test, we computed the optical depth along every coordinate diagonal. In the second, we computed the optical
depth by finding the path of smallest optical depth among all paths leading from the current point to a domain boundary (approximated
as all paths formed of cell-to-cell piecewise segments and connecting the current point to the outer boundary). We
do this by computing the optical depth at a cell center $\tau_c$ from the optical depth at the center of neighboring cells $\tau_i$ 
and the optical depth along the line connecting the center of the cell to the center of the neighboring cells $\tau_{c-i}$, 
i.e.
\beq
\tau_c = \min_i(\tau_i + \tau_{c-i}).\label{eq:tauNeilsen}
\eeq 
Indeed, if the $\tau_{c-i}$ are the optical depths along the paths through which neutrinos can escaped the most easily
from the neighboring cells `$i$', then Eq.~(\ref{eq:tauNeilsen}) returns the optical depth along the path through
which neutrinos can the most easily escape from the current cell `$c$'.
We thus only need to iterate Eq.~(\ref{eq:tauNeilsen}) until $\tau_c$ converges at all
points -- which occurs rapidly if $\tau_c$ is initialized to its value at the previous time step.
This method is inspired by the recent
work of Perego et al.~\cite{2014arXiv1403.1297P} and Neilsen et al.~\cite{2014arXiv1403.3680N}, and closely follows the algorithm presented
in~\cite{2014arXiv1403.3680N}. Both tests lead to changes of $\sim 20\%$ in the neutrino luminosities, thus showing that the exact method
used to compute the optical depth is a fairly minor source of error compared to the systematic error due to the use of a leakage scheme. 

As a last test of our code for simulations with $M_{\rm BH}=7M_\odot$, 
we consider the effect of the treatment of the low-density region. This region is dynamically unimportant, but could conceivably 
affect the neutrino luminosities if it develops a very hot atmosphere. We also study the impact of the black hole excision 
by using linear extrapolation of the density and velocities to the faces of our grid located at the border 
of the excised region instead of just copying the value of the fields from the nearest cell center, as well as by limiting the temperature
of the fluid on the excision boundary to $T=10\,{\rm MeV}$. The resulting changes in the
post-merger properties of the system are well below our estimates of the numerical error due to the finite resolution of our grid. 
%(and the largest changes are observed
%in the high-velocity tail of the velocity distribution of the ejecta and in the magnitude of the neutrino luminosity in the first 
%$2{\rm ms}-3{\rm ms}$ after merger).

\begin{figure*}
%{\includegraphics[width=0.9\columnwidth,bb=0 6 142 127]{LS220DisruptKm}}
%\includegraphics[width=0.9\columnwidth,bb=0 4 145 133]{G2DisruptKm}
\includegraphics[width=0.9\columnwidth]{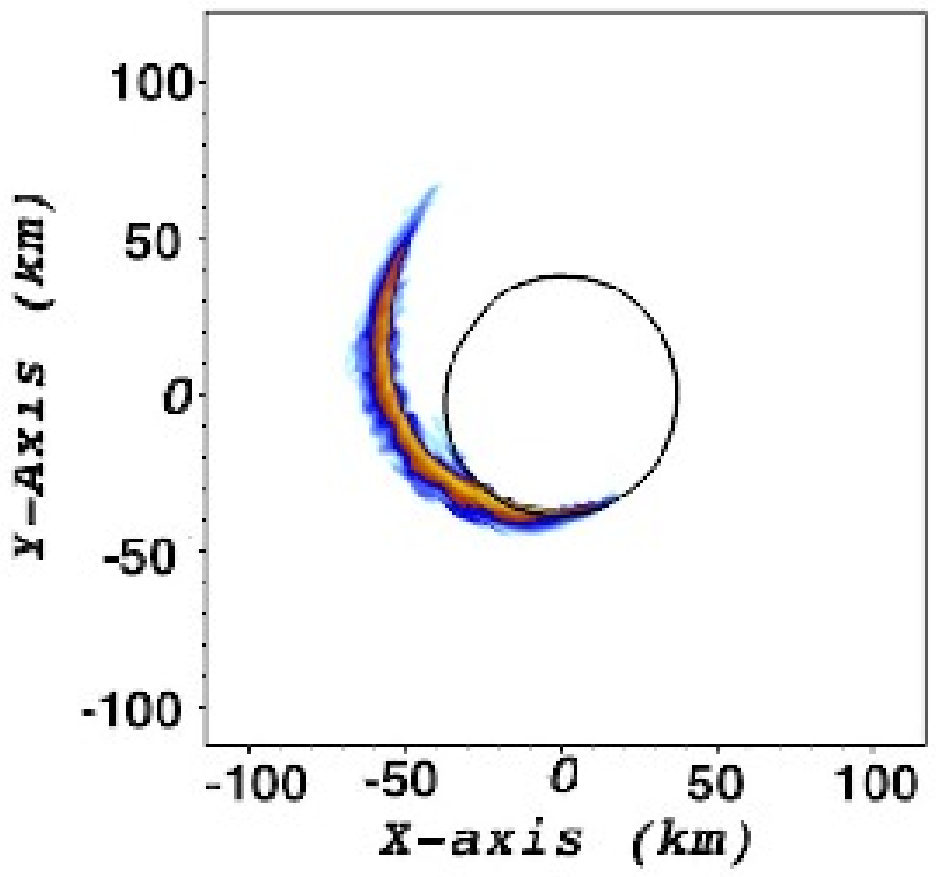}
\includegraphics[width=0.9\columnwidth]{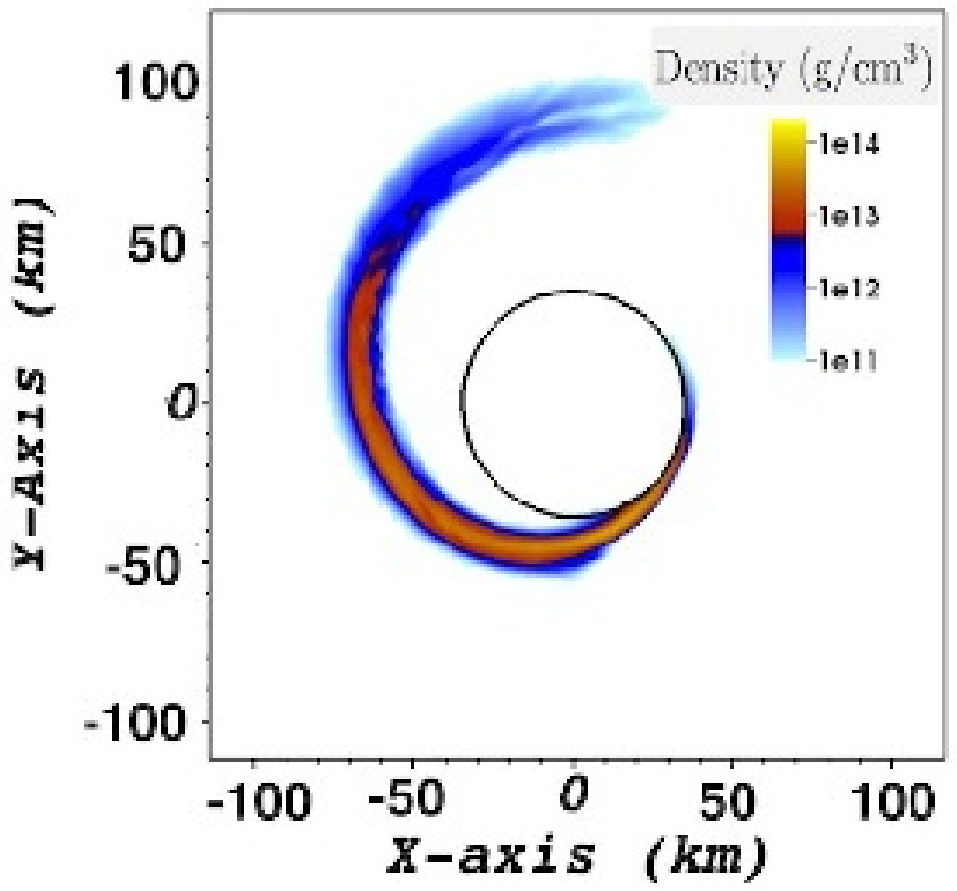}
\caption{Comparison of the distribution of matter during disruption for the LS220 equation of state used in this work ({\it left}, simulation M14-10-S8), 
and a $\Gamma=2$ equation of state ({\it right}, simulation Q7S9-R12i0 from~\cite{Foucart:2013a}). Snapshots taken in the
equatorial plane at a time at which $0.25M_\odot$ of
material remains outside of the black hole. Both simulations lead to $\sim 0.1M_\odot$
of material remaining outside of the black hole at late times - yet the distribution of that material between bound and unbound material
and the properties of the tidal tail are very different. In each case, the thick black line shows the location of the apparent horizon of the
black hole.}
\label{fig:G2vsLS220}
\end{figure*}

We now turn our attention to the higher mass ratio configurations. These simulations are significantly more difficult to resolve
due to a number of issues:
\begin{itemize}
\item The disruption of the neutron star occurs very close to the black hole: except for simulation M12-10-S9, the tidal tail begins to
form as high-density material has already begun to cross the apparent horizon of the black hole. But the accuracy of our code
is lower close to the apparent horizon, due to the use of one-sided stencils close to the excised region inside of the horizon. 
Additionally, stricter
control of the velocities and temperature within the low-density regions is necessary for stability in that excised region. Hence, properly
resolving the disruption of the neutron star requires higher resolution in such cases. 
\item Compared to previous simulations using a Gamma-law ($\Gamma=2$) equation of state~\cite{Foucart:2013a}, 
simulations using the LS220 equation of state show important
differences in the qualitative features of the tidal disruption. Initially, the neutron star remains more compact with the LS220 equation
of state, and most of the expansion of the neutron star occurs later in the plunge (see Fig.~\ref{fig:G2vsLS220}), but more rapidly. This makes
the disruption of the neutron star more difficult to resolve with the nuclear-theory based equation of state. This is ultimately
a consequence of the large difference between the stiffness of the LS220 equation of state at high and low densities (with respect
to nuclear saturation density).
\item Once a tidal tail forms, tidal compression is very efficient and forces the tail to be much thinner than for a $\Gamma=2$ equation of state
(see Fig.~\ref{fig:G2vsLS220}). This is, again, a consequence of the properties of the equation of state at low densities. Because disruption
occurs as the neutron star is already plunging into the black hole, the effects of tidal compression can be significant. We find that the
tail has a typical coordinate width (and vertical height) of only a few kilometers
(note that Fig.~\ref{fig:G2vsLS220} uses a logarithmic scale for the density). 
This should be compared with the radius of the apparent horizon of the black hole, which in our gauge is $\sim 30\,{\rm km}$.
Accordingly, the typically resolutions used in the simulation of NSBH mergers are insufficient in this case.
Indeed, simulations using fixed mesh refinement typically use $20\textrm{--}30$ grid points across the radius of the 
black hole (see e.g.~\cite{Kyutoku:2011vz}), while our finite difference grid generally has a higher 
resolution at the beginning of the disruption ($\sim R_{\rm NS}/50\sim 0.2\,{\rm km}$), 
but rapidly becomes coarser as we progressively expand the grid in order to follow the expansion of the tidal tail. Additionally,
our simulations suffer from lower accuracy in the immediate neighborhood of the excised region, as opposed to codes which do not excise
the inside of the black hole and can use high-order finite difference stencils crossing the horizon. In SpEC, and for large black hole
spins, the excised region can be very close to the horizon of the black hole~\cite{Hemberger:2012jz}, and thus some regions
outside of the horizon have to use low-order stencils including points on the boundary of the evolved computational domain
where errors in the fluid properties are larger. 
\end{itemize}
Accordingly, as soon as the resolution of our finite difference grid drops below $\Delta x\approx 0.5\,{\rm km}\textrm{--}1\,{\rm km}$, simulations
with $M_{\rm BH}=10M_\odot$ and the LS220 equation of state 
suffer great losses of accuracy. This typically takes the form of a rapid spurious acceleration of the tail material
within the unresolved region, which then becomes unbound with large asymptotic velocities ($v/c \sim 0.6\textrm{--}0.9$). To resolve the tidal
tail during this phase requires resolution that our code cannot maintain while following the unbound ejecta and bound tidal tail 
far away from the black hole. One solution is to use mesh refinement for the finite difference grid.
Although we have implemented a fixed mesh refinement algorithm for our finite difference grid, the cost
of interpolating between the different evolution grids currently makes it impractical to use more than 2-3 levels of refinement,
and thus to resolve accurately the material close to the black hole while following the ejecta for a few milliseconds,
especially for high mass ratio simulations which require a small grid spacing on both our finite difference and pseudospectral
grids. 
Instead. we choose to continue simulations M12-10-S8 and M12-10-S9 through disruption while maintaining a resolution in
the region close to the black hole of $\Delta x<0.7\,{\rm km}$ (and vertically $\Delta x = 0.2\,{\rm km}$) for M12-10-S9, with 2 levels
of refinement, and $\Delta x < 0.4\,{\rm km}$ (and vertically, $\Delta x = 0.1\,{\rm km}$) for the more challenging M12-10-S8 
%(for the lower spin configuration, the disruption occurs closer to the apparent horizon of the black hole)
, with 3 level of refinements.
Each refinement level is a box using $140^3$ grid points, and the grid spacing is multiplied by a factor of 2 between levels. 
Both of those simulations confirmed that the rapid acceleration observed at lower resolution is indeed a numerical artifact. 
The higher resolution simulations eject $\sim 0.1M_\odot\textrm{--}0.15M_\odot$ of material at velocities $v/c \lesssim 0.5$ . 
Using  a fixed mesh refinement with a resolution similar to the one used in simulation M12-10-S9 proved to be insufficient to resolve
the configurations for which the disruption occurs as the neutron star crosses the apparent horizon of the black hole
(M12-10-S8, M14-10-S8, M14-10-S9). Clearly, with these grids covering such a small region around the black hole 
(up to $\sim 5M_{\rm BH}\sim 75\,{\rm km}$), we can only measure the ejected mass and the mass remaining outside of the black hole (and the latter
will only be a rough estimate). We cannot follow the evolution of the disk as it expands and is impacted by fall-back from the 
tidal tail. Hence, for the high mass ratio simulations with fixed mesh refinement (M12-10-S8, M12-10-S9), 
we stop the evolution when an accretion disk forms.
Similarly, for the high mass ratio simulations without mesh refinement (M14-10-S8, M14-10-S9), we only obtain estimates of
the ejected mass and of the mass remaining outside of the black hole: we cannot reliably continue the evolution once the 
resolution drops below $\Delta x\approx 0.5\,{\rm km}\textrm{--}1\,{\rm km}$.

\subsection{Tracer particles}
\label{sec:tracers}

A major motivation for the simulations presented here is to characterize the
unbound outflow:  its mass, asymptotic velocity distribution, composition, and
entropy. In a stationary spacetime and in the absence of pressure forces,
a fluid element with 4-velocity $u^{\alpha}$ is unbound if it has
positive specific energy $E=-u_t-1$ (under those conditions,
$E$ is constant). In our simulations, we consider a fluid element as unbound if
$E>0$ as it leaves the grid -- even though it has a finite pressure 
and the metric is not yet completely stationary. 
Thus, another potential source of error in our conclusions comes from the finite
outer boundary of our fluid evolution grid.

To test that the outflow reaches its asymptotic state before exiting the
simulation, we evolve one hundred Lagrangian tracer particles in the
weakly bound and unbound flow.  That is, the position of tracer particle
A is evolved according to the local 3-velocity:  $dx_A^i/dt = v^i(x_A^i)$,
and fluid quantities at tracer positions are monitored by interpolating
from the fluid evolution grid.

\begin{figure}
\flushleft
\includegraphics[width=0.95\columnwidth]{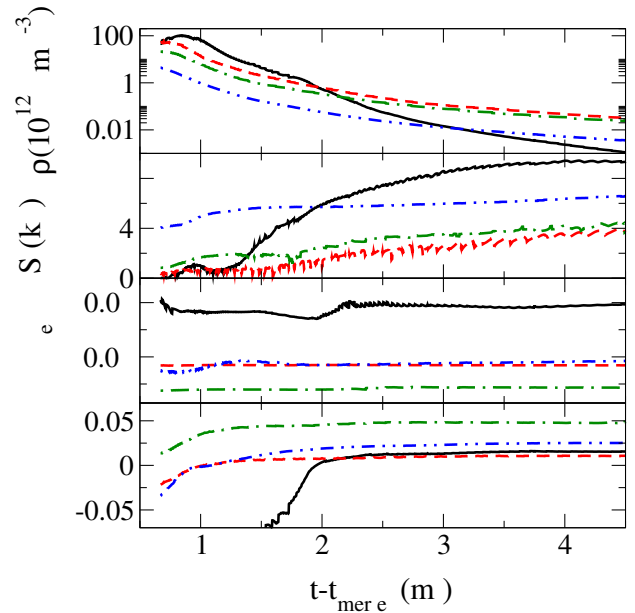}
\caption{The evolution of four representative tracer particles
  representing unbound fluid elements.  We plot the baryon density $\rho$,
  the specific entropy $S$, the electron fraction $Y_e$, and the
  specific energy $E\equiv -u_t-1$.  The small glitches in the
  entropy evolution are artifacts of stencil changes with the
  low-order interpolator used for Lagrangian output.
  Note, the tracer particle denoted by the solid black line
  begins closest to the black hole.}
\label{fig:Tracer}
\end{figure}

In Fig.~\ref{fig:Tracer}, we plot some fluid quantities for four
representative randomly chosen unbound particles in the ejecta of
case M12-7-S9.  We see that the
energies stabilize after only about two milliseconds of post-merger
evolution.  The electron fraction $Y_e$ becomes stationary the soonest
because the tail is too cold for strong charged-current interactions
that would change $Y_e$. 
Entropy also levels off, albeit more slowly.  
Over the $4\,{\rm ms}$ of evolution, the
density decreases by about three orders of magnitude.

We are thus confident that near-merger gravitational fluctuations,
pressure forces, and neutrino emission do not significantly contaminate
our predictions about the outflows.  It is important to note that the
entropy and energy may change significantly outside the evolved region
because of recombination and nucleosynthesis in the decompressing
material.  Charged current heating from the background neutrino
field, which we do not include in our simulations, can also change the
composition of the ejecta, particularly
for material unbound by disk winds (as opposed to the material
ejected during the disruption of the neutron star).
The late-time, large-distance evolution of the ejecta is
a matter of ongoing study. 

\section{Disruption, Disk Formation and Outflows}
\label{sec:dynamics}

\subsection{Disruption of the neutron star}
\label{sec:merger}

Let us now consider the results of these simulations, starting with the disruption and merger dynamics.
At the time of tidal disruption, the results are largely unaffected by neutrinos (which act
over timescales $\sim 10\,{\rm ms}$, while the merger takes $\sim 1\,{\rm ms}$). Instead, the main
components required to properly model the system are the size and structure of the neutron star,
and general relativity. 

We choose the initial parameters of these simulations so that the neutron star is disrupted, thus allowing
the formation of an accretion disk and/or the ejection of material. This is indeed what we observe.  
The mass remaining outside of the black hole after disruption (including unbound material) and
the final black hole mass and spin are in agreement with previous numerical studies
using simpler equations of state,
either in the same range of parameters~\cite{Foucart:2013a}, or for slightly lower black hole 
spins~\cite{Kyutoku:2011vz}. They are also largely consistent
with semi-analytical models mostly based on lower mass simulations with simpler equations
of state~\cite{Foucart2012,Pannarale:2012}. 
However, going beyond these global properties of the post-merger remnant, we observe
differences in the qualitative features of the disruption phase, modifying the 
distribution of the material remaining outside of the black hole between unbound material, bound
tail and accretion disk (summarized in Table~\ref{tab:remnant}).

\begin{table*}
\caption{
Properties of the final remnant. $M_{\rm BH}^f$ is the mass of the black hole at the end of the simulation, $\chi_{\rm BH}^f$
is its dimensionless spin, $v_{\rm kick,GW}$ is the predicted recoil velocity due to gravitational wave emission (i.e. not taking into
account the larger recoil from the asymmetric ejection of material), while $v_{\rm recoil}^{BH}$ is the measured coordinate velocity
of the black hole after merger. $M_{\rm out}^{5{\rm ms}}$ is the baryon mass outside of the black hole
$5\,{\rm ms}$ after merger (including unbound material). $M_{\rm disk}^{20{\rm ms}}$, $M_{\rm tail,bound}^{20{\rm ms}}$ and $M_{\rm ej}$ are the baryon mass of the
disk, bound tail material, and unbound material $20\,{\rm ms}$ after merger. By conservation of baryon number, 
$M_{\rm out}^{5{\rm ms}}-M_{\rm disk}^{20{\rm ms}}-M_{\rm tail,bound}^{20{\rm ms}}-M_{\rm ej}$ is the mass accreted onto the black hole
in the $15\,{\rm ms}$ between the two sets of measurements. $E_{\rm ej}$ and $\langle v_{\rm ej}\rangle$
are the kinetic energy and average asymptotic velocity of the unbound material. Fluid quantities for the 
higher mass ratio simulations (second half of the table) generally have larger uncertainties, as discussed 
in Sec.~\ref{sec:errors}, and are only provided when they can be reliably extracted from the simulation.
}
\label{tab:remnant}
\begin{tabular}{|c||c|c|c|c|c|c|c|c|c|c|}
\hline
Name & $M_{\rm BH}^f$ & $\chi_{\rm BH}^f$ & $v_{\rm kick}^{\rm GW}$ (km/s) & $v_{\rm recoil}^{\rm BH}$ (km/s)& $M_{\rm out}^{5{\rm ms}}$ & $M_{\rm disk}^{20{\rm ms}}$ & $M_{\rm tail,bound}^{20{\rm ms}}$ & $M_{\rm ej}$ & $E_{\rm ej}(10^{51}{\rm erg})$ & $\langle v_{\rm ej} \rangle /c$ \\
\hline
M12-7-S8 & $7.79M_\odot$ & 0.85 & 60 & 675 & $0.31M_\odot$ & $0.09M_\odot$ & $0.04M_\odot$ & $0.14M_\odot$ & 8.5 & 0.24 \\ 
M12-7-S9 & $7.74M_\odot$ & 0.91 & 69 & 874 & $0.37M_\odot$ & $0.13M_\odot$ & $0.03M_\odot$ & $0.16M_\odot$ & 11 & 0.25 \\
M14-7-S7 & $8.07M_\odot$ & 0.81 & 79 & 151 & $0.15M_\odot$ & $0.04M_\odot$ & $0.05M_\odot$ & $0.04M_\odot$ & 1.7 & 0.20 \\
M14-7-S8 & $8.00M_\odot$ & 0.87 & 76 & 185 & $0.25M_\odot$ & $0.08M_\odot$ & $0.05M_\odot$ & $0.06M_\odot$ & 2.0 & 0.18 \\
M14-7-S9 & $7.95M_\odot$ & 0.92 & 94 & 345 & $0.35M_\odot$ & $0.14M_\odot$ & $0.04M_\odot$ & $0.07M_\odot$ & 2.5 & 0.18 \\
\hline
M12-10-S8 & $10.75M_\odot$ & 0.83 & 45 & -- & $\sim 0.25M_\odot$ & -- & -- & $\sim (0.10\textrm{--}0.15)M_\odot$ & -- & --\\
M12-10-S9 & $10.58M_\odot$ & 0.89 & 49 & -- & $\sim 0.40M_\odot$ & -- & -- & $\sim (0.10\textrm{--}0.15)M_\odot$ & $\sim 10\textrm{--}20$ & $\sim 0.3-0.35$ \\
M14-10-S8 & $11.00M_\odot$ & 0.85 & 22 & -- & $\sim 0.10M_\odot$ & -- & -- & $\sim (0.05\textrm{--}0.10)M_\odot$ & -- & --\\
M14-10-S9 & $10.70M_\odot$ & 0.90 & 54 & -- & $\sim 0.30M_\odot$ & -- & -- & $\sim (0.10\textrm{--}0.15)M_\odot$ & -- & --\\
\hline
\end{tabular}
\end{table*}

The systems with the more symmetric mass ratio ($q=5$, with $M_{\rm NS}=1.4M_\odot$ and $M_{\rm BH}=7M_\odot$) 
provide the most traditional results: between $0.15M_\odot$ and
$0.35M_\odot$ of material remains outside of the black hole $5\,{\rm ms}$ after merger. The higher black hole spins
naturally lead to more massive remnants, as expected from the fact that the innermost stable circular orbit is at 
a lower radius for prograde orbits around a more rapidly spinning black hole. We find that the system ejects 
$0.04M_\odot\textrm{--}0.07M_\odot$ of material, which appears compatible with the range of ejected mass listed for 
NSBH binaries at $\chi_{\rm BH}=0.75$ by Hotokezaka et al.~\cite{2013ApJ...778L..16H}: 
although~\cite{2013ApJ...778L..16H} does not list values for individual systems,
they find a range $M_{\rm ej}\sim 0.02M_\odot\textrm{--}0.04M_\odot$ for the neutron star closest in size to the $1.4M_\odot$ neutron star
used in this work and mass ratios $q=3\textrm{--}7$. The higher masses found here can easily be attributed to larger black
hole spins. The average velocity of the ejecta is $\langle v \rangle \sim 0.2c$,
where the average $\langle X \rangle$ of a quantity $X$ on a spatial slice $\Sigma$ is here defined as
\beq
\langle X \rangle = \frac{\int_\Sigma \rho X W\sqrt{g} dV}{\int_\Sigma \rho W\sqrt{g} dV},
\eeq
$\rho$ is the baryon density of the fluid, $W$ its Lorentz factor, 
and $g$ is the determinant of the 3-metric on slice $\Sigma$. This is slightly lower than
in~\cite{2013ApJ...778L..16H} ($\langle v \rangle \sim 0.25c\textrm{--}0.29 c$).
As for the bound material, $20\,{\rm ms}$ after merger $0.04M_\odot\textrm{--}0.14M_\odot$ has formed a hot accretion disk cooled 
by neutrinos (discussed in Sections~\ref{sec:disk}-\ref{sec:neutrino}), $0.04M_\odot-0.05M_\odot$ has left the grid on
highly eccentric bound trajectories, while the rest ($0.02M_\odot\textrm{--}0.10M_\odot$) has accreted onto the black hole.
Extrapolating the remnant mass to lower spins indicate that the neutron star will disrupt for $\chi_{\rm BH}\gtrsim 0.55$,
as predicted from simulations with simpler equations of state~\cite{Foucart2012}. 

\begin{figure}
\includegraphics[width=0.95\columnwidth]{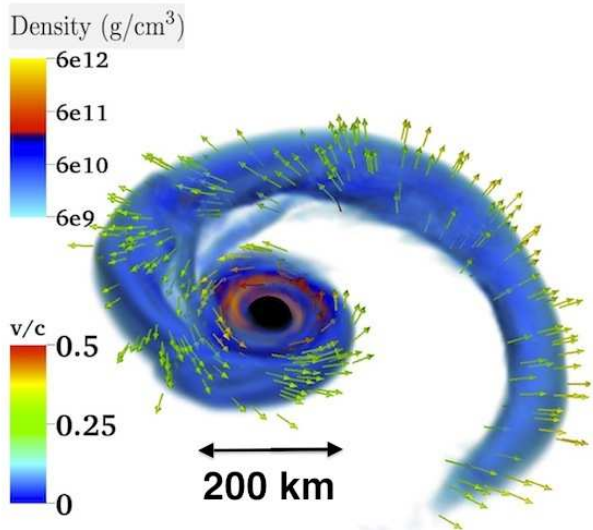}
\caption{Matter distribution during the disruption of the neutron star for case M12-7-S8. About half of the remnant material is unbound,
while a relatively low mass hot disk forms.}
\label{fig:VolEj}
\end{figure}

At higher mass ratios, we generally find that a larger amount of material is ejected.
Our two simulations with $M_{\rm NS}=1.2M_\odot$ and $M_{\rm BH}=7M_\odot$ find about the same amount of material
in the disk and bound tail as for the $q=5$ cases ($M_{\rm disk}\sim0.1M_\odot$, $M_{\rm tail}\sim 0.05M_\odot$),
but about $0.15M_\odot$ is ejected at speeds $\langle v \rangle \sim 0.25c$ (see Fig.~\ref{fig:VolEj}).
The closest results to compare to are again those of Hotokezaka et al.~\cite{2013ApJ...778L..16H}, for the H4 equation
of state, which has a compactness $C_{\rm NS}=0.147$ for a mass $M_{\rm NS}=1.35$. 
Hotokezaka et al.~\cite{2013ApJ...778L..16H} find $M_{\rm ej}=0.04M_\odot\textrm{--}0.05M_\odot$ for $q=3-7$ and $\chi_{\rm BH}=0.75$,
and similar average velocities. Even considering the different spins used and the expected error bars, 
the more massive ejecta
found here appear to be an indication of a dependence of the ejected mass on the internal structure of the neutron star:
from our results, we would predict $M_{\rm ej}\sim 0.13M_\odot$ for $q=5.8$ and $\chi_{\rm BH}=0.75$. The difference with
the results of Hotokezaka et al.~\cite{2013ApJ...778L..16H} is slightly
out of the $60\%$ relative error that we consider to be a strict upper bound on the error in our measurement of $M_{\rm ej}$,
and which is most likely a significant overestimate of that error.
The most likely reason for an equation of state dependence of the ejected mass at higher mass ratios is that 
the disruption of the neutron star then occurs as the neutron star is plunging into the black hole. 
The mass and properties of the ejected material then
depend not only on the separation at which disruption occurs, but also on the time-dependent response of the neutron
star to the tidal disruption. For the same reason, the properties of the post-merger remnant 
also have a steeper dependence in the mass and spin of the black hole for more massive black holes.

As discussed in Sec.~\ref{sec:errors}, for the simulations with $M_{\rm BH}=10M_\odot$ we can only
resolve the disruption of the neutron star and rapid accretion onto the black hole if we use a very high
resolution grid covering only a small area around the black hole. This is due largely to qualitative differences
in the disruption process: the neutron star disrupts as high-density material has already begun to cross the
apparent horizon, and the entire disruption and tail formation process occurs within a distance of
about $3 M_{\rm BH}\sim 45\,{\rm km}$
of the black hole center. All material surviving the merger is on highly eccentric orbits, and the tidal
tail experiences strong tidal compression in the directions in which the tidal field of the black hole
causes trajectories to converge. The tidal tail is reduced to a thin stream of matter only a few kilometers
wide. Not surprisingly, in this regime the result of the merger is very sensitive to the parameters of the binary:
for example, a small change in the spin of the black hole (e.g.\ $\delta \chi_{\rm BH}\sim 0.1$, as in our simulations)
drastically changes the amount of material which remains
outside of the black hole after disruption. Changes in the stiffness of the equation of state, which affect the distribution
of matter during disruption, also become even more important than in the previously discussed configurations
(see Fig.~\ref{fig:G2vsLS220} for a comparison with a similar simulation with a $\Gamma=2$ equation of state).
Surprisingly, even in this case, the total amount of material surviving disruption remains similar for both
the $\Gamma=2$ and LS220 equations of state. We can for example compare simulation M14-10-S9 of this work, 
for which about $0.3M_\odot$
remains outside the black hole after merger, and simulation R13i0 of Foucart et al. 2013~\cite{Foucart:2013a},
with the same black hole parameters and a similar neutron star radius $R_{\rm NS}=13.3\,{\rm km}$, in which
$0.31M_\odot$ of material remains. But more material is unbound during merger for the LS220 equation of 
state: $M_{\rm ej}\gtrsim 0.1M_\odot$ here, while $M_{\rm ej}\sim 0.05M_\odot$ in 
Foucart et al. 2013~\cite{Foucart:2013a}. And the disk is initially
less massive here: half of the material promptly formed a disk in~\cite{Foucart:2013a}, nearly all
of the material is on highly eccentric orbits here. The long term evolution of the disk will thus be more 
significantly affected by the fall-back of tail material than expected from the simulations with a
$\Gamma=2$ equation of state. Compared to the analytical prediction of~\cite{Foucart2012}, there is an
excess of material surviving the disruption, which is what we usually find for high spin black holes 
and high mass ratio. But the estimate that disruption will only occur for $\chi_{\rm BH}\gtrsim 0.65$ (resp.\ 
$\chi_{\rm BH}\gtrsim 0.75$) for $M_{\rm NS}=1.4M_\odot$ (resp.\ $1.2M_\odot$) appears
to be accurate -- although it is of course dangerous to draw such a conclusion by
extrapolating from only 2 simulations with inaccurate estimates of the remnant mass.

\begin{figure}
\flushleft
\includegraphics[width=0.85\columnwidth]{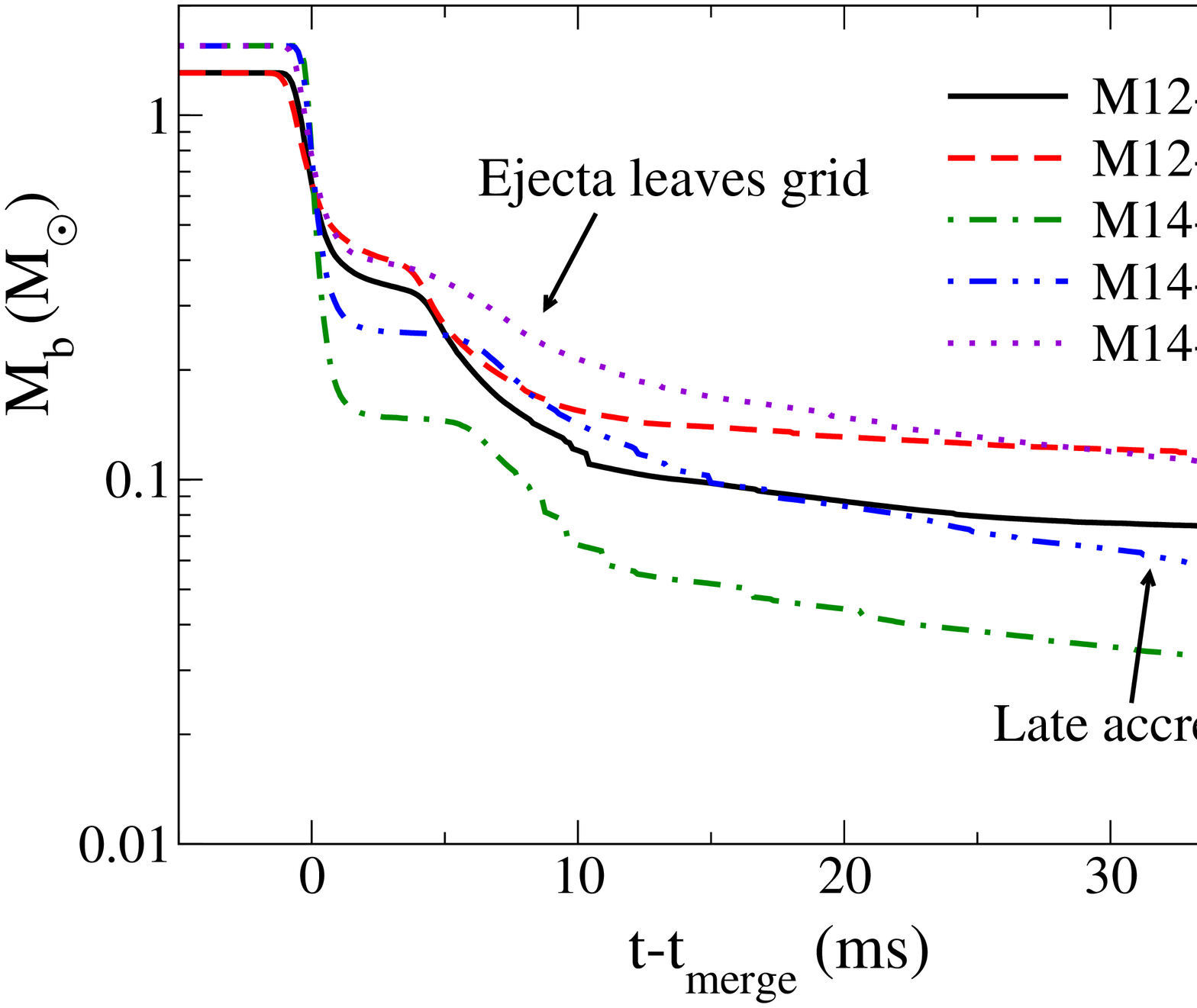}
\caption{Baryon mass left on the numerical grid as a function of time,
for all simulations with $M_{\rm BH}=7M_\odot$.}
\label{fig:BaryonMass}
\end{figure}

All of the simulations presented here thus appear to have post-merger remnants with 
large amounts of both bound and unbound material, providing promising setups for potential electromagnetic
signals. The typical timescales for the merger can be observed on Fig.~\ref{fig:BaryonMass}, which shows
the baryon mass remaining on the grid at any given time
for all simulations with $M_{\rm BH}=7M_\odot$.
The disruption of the neutron star and rapid accretion of material onto the black hole occur over $\sim 1\,{\rm ms}$.
Then accretion slows down and the disk evolves over timescales of $\sim 0.1{\rm s}$. The rapid variation in the mass 
present on the grid around $5\,{\rm ms}\textrm{--}10\,{\rm ms}$ after merger is due to the unbound material leaving the grid (and discrete
jumps are due to modifications of the location of the outer boundary of the grid as the algorithm in charge of deciding which
regions of the spacetime to cover abandons low-density regions containing unbound or marginally bound material).
Not shown are the simulations for $M_{\rm BH}=10M_\odot$, for which most of the material escaping early accretion onto the black hole
rapidly leaves the grid. In the following sections, we will now look in more detail into the properties of the unbound
material and the evolution of the disk.

\subsection{Properties of the outflows}
\label{sec:ejecta}

Given the large mass of material ejected by these mergers, NSBH binaries with $M_{\rm BH}\sim 7M_\odot\textrm{--}10M_\odot$ appear to be
prime candidates to produce post-merger electromagnetic counterparts, and large quantities of r-process elements
(see Table~\ref{tab:remnant}).
Although the mass and velocity of these outflows vary from case to case, other properties of the ejecta are more consistent. Since the material
is ejected before neutrino emission has had a chance to modify its composition, it has a low electron fraction ($Y_e<0.1$). 
The neutron star matter initially has a low
entropy per baryon, and is not significantly heated during the tidal compression of the tail and the ejection of material:
the entropy per baryon of the ejecta for the simulations with  $M_{\rm BH}=7M_\odot$ is 
$\langle S \rangle/k_B\sim 4-5$, and it is only slightly higher for simulations with $M_{\rm BH}=10M_\odot$ 
($\langle S\rangle /k_B\sim 5-7$). 
The outflow is expected to robustly undergo r-process 
nucleosynthesis, with the production of material with $A>120$ insensitive to variations in the properties of the outflow within the 
range of values observed here~\cite{Roberts2011}. Accordingly, all of the
simulations considered here are promising setups for the production of infrared transients~\cite{Roberts2011,metzger:11} and 
heavy r-process elements.

\begin{figure}
\flushleft
\includegraphics[width=0.85\columnwidth]{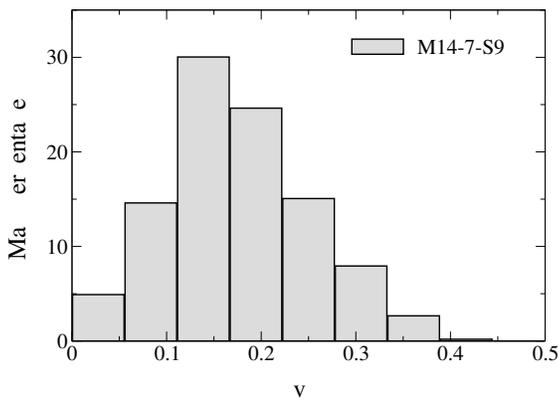}
\caption{Distribution of asymptotic velocities for the unbound material in simulation M14-7-S9.
Other simulations with $M_{\rm BH}=7M_\odot$ have similar distributions, with a slightly higher
peak value when $M_{\rm NS}=1.2M_\odot$.}
\label{fig:Vbin}
\end{figure}

The asymptotic velocity distribution of the ejecta (its expected velocity at infinite distance from the black hole)
is shown in Fig.~\ref{fig:Vbin} for simulation M14-7-S9.
For all configurations with $M_{\rm BH}=7M_\odot$, we find similar distributions spanning $0<v<0.5c$,
with slight variations in the location of the peak. A more asymmetric mass ratio generally
results in higher asymptotic velocities for the ejecta. This additional acceleration is presumably due to the fact that,
since the disruption occurs on an eccentric orbit passing close to the black hole, the fluid can be
efficiently accelerated by the rapid compression and decompression of the tidal tail under the influence
of the tidal field of the black hole (see also Sec.~\ref{sec:merger}). This effect is specific to high mass ratio binaries, for which the
disruption occurs as the neutron star plunges into the black hole and the properties of the disrupted
material are sensitive to the equation of state (see Fig.~\ref{fig:G2vsLS220}).

As pointed out in Kyutoku et al.\ 2013~\cite{2013PhRvD..88d1503K}, the ejection of unbound
material is not isotropic.
In our simulations, the ejecta is confined within $\sim 20^\circ$ of the equatorial plane, and covers an azimuthal angle
$\Delta \phi \sim \pi$ (see Fig.~\ref{fig:VolEj}). An important consequence of this anisotropy is that a kick is imparted
onto the black hole with $v_{\rm kick,ej}\sim M_{\rm ej}v_{\rm ej}/M_{\rm BH}$.
%For simulation M14-10-S9, for which we follow tracer particles, we measure the linear momentum carried 
For simulation M12-7-S9, for which we follow tracer particles, we measure the linear momentum carried 
away by the ejecta and find
\beq
v_{\rm kick,ej}\sim 0.5 \frac{M_{\rm ej}}{M_{\rm BH}}\langle v\rangle _{\rm ej} \sim 770\,{\rm km/s}\,\,\label{eq:vkick}.
\eeq
The factor of $0.5$ comes from the fact that not all of the ejecta is ejected in the same direction.
We estimate it from the direction of $v^i$ as the tracer particles cross spheres of constant radius.
Even considering the uncertainties in our measurements of $M_{\rm ej}$ and $\langle v\rangle _{\rm ej}$, this
is clearly larger than the kick velocity due to gravitational wave emission 
(see Table~\ref{tab:remnant}): our simulations predict 
$v_{\rm kick,ej}\sim 150\,{\rm km/s}\textrm{--}800\,{\rm km/s}$ but $v_{\rm kick,GW}\sim 20\,{\rm km/s}\textrm{--}100\,{\rm km/s}$.
These velocities play an important role when assessing whether globular clusters can retain 
a black hole after it merges with a neutron star, and in the determination of the
rate of NSBH mergers occurring in those clusters~\cite{2013MNRAS.428.3618C}. At the high
end of this velocity range, the kick could even be above the escape velocity of small galaxies. 

The amplitude of this kick is qualitatively confirmed by measurements of the coordinate velocity
of the black hole, defined as its average velocity on the grid between $5\,{\rm ms}$ and $15\,{\rm ms}$
after merger. Although coordinate velocities are gauge dependent and should thus be taken
with a certain skepticism, the values measured here are compatible with the predictions
from Eq.~(\ref{eq:vkick}) while being much larger than the predicted recoil velocities
due to asymmetric emission of gravitational waves (see Table~\ref{tab:remnant}).

\subsection{Disk evolution}
\label{sec:disk}

Constraining the formation and the long-term evolution of the accretion disks resulting from NSBH mergers is an important step towards
assessing their potential as short gamma-ray burst progenitors. Those disks are also likely to power outflows (e.g.\ via neutrino-driven 
or magnetically-driven winds, or recombination of $\alpha$-particles), causing optical and radio signals. 

\begin{figure}
\flushleft
\includegraphics[width=0.85\columnwidth]{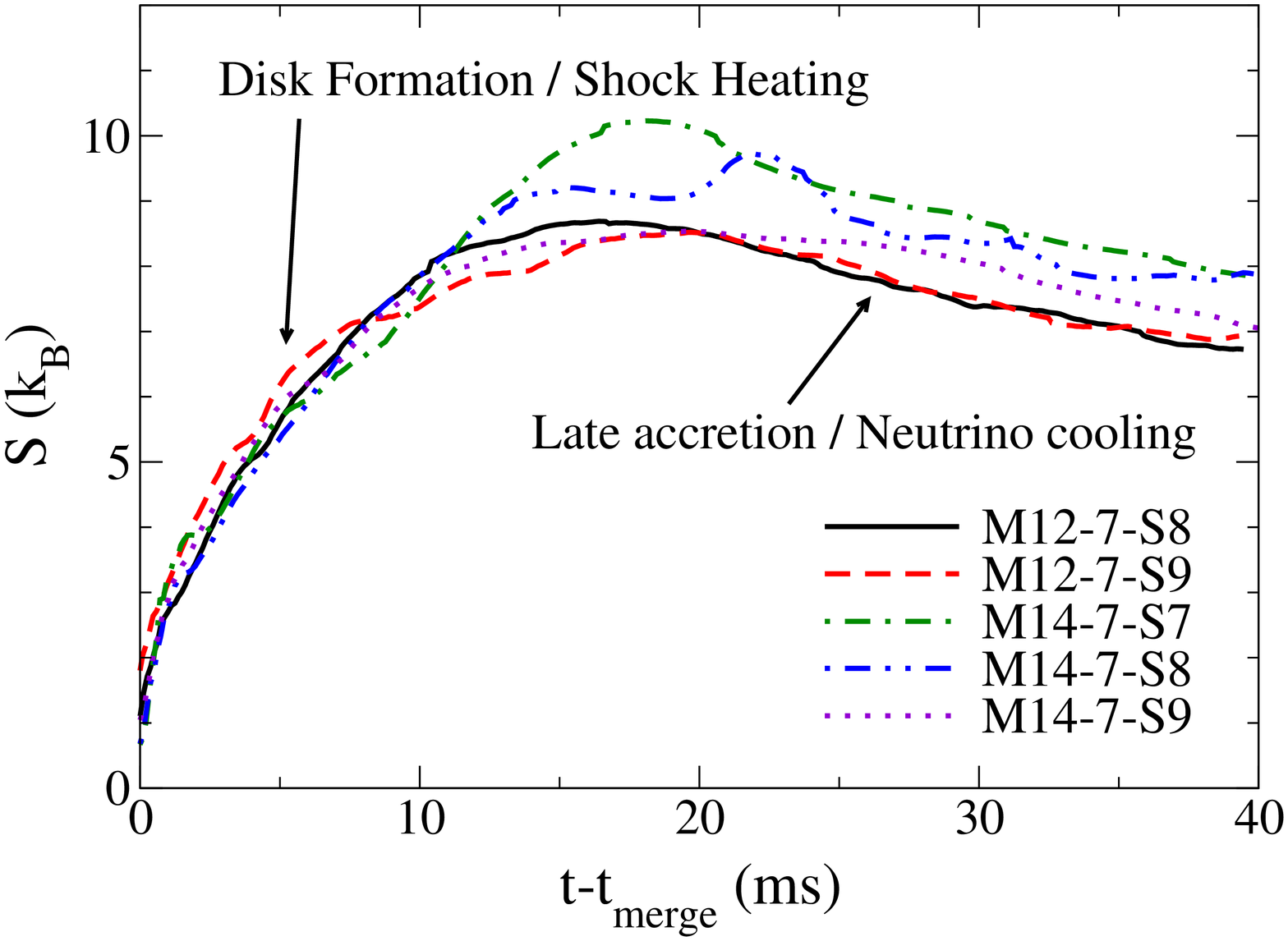}
\caption{Average entropy per baryon on the computational grid as a function of time,
for all simulations with $M_{\rm BH}=7M_\odot$.}
\label{fig:Entropy}
\end{figure}

\begin{figure}
\flushleft
\includegraphics[width=0.85\columnwidth]{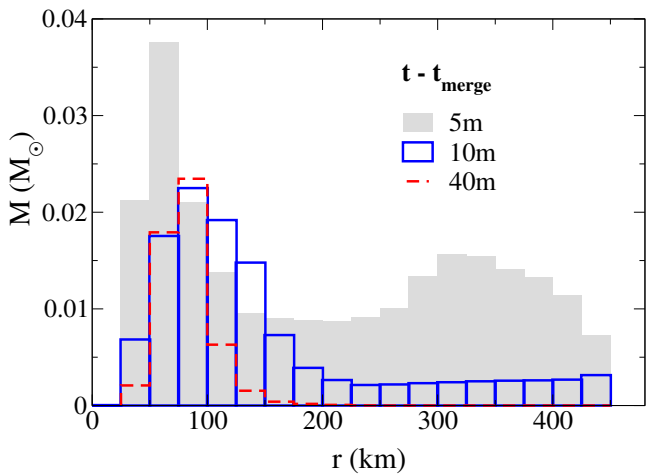}
\caption{Distribution of matter around the black hole measured $5\,{\rm ms}$, $10\,{\rm ms}$ and $40\,{\rm ms}$
after merger, for simulation M14-7-S8.}
\label{fig:RhoVsR}
\end{figure}

\begin{figure}
\flushleft
\includegraphics[width=0.85\columnwidth]{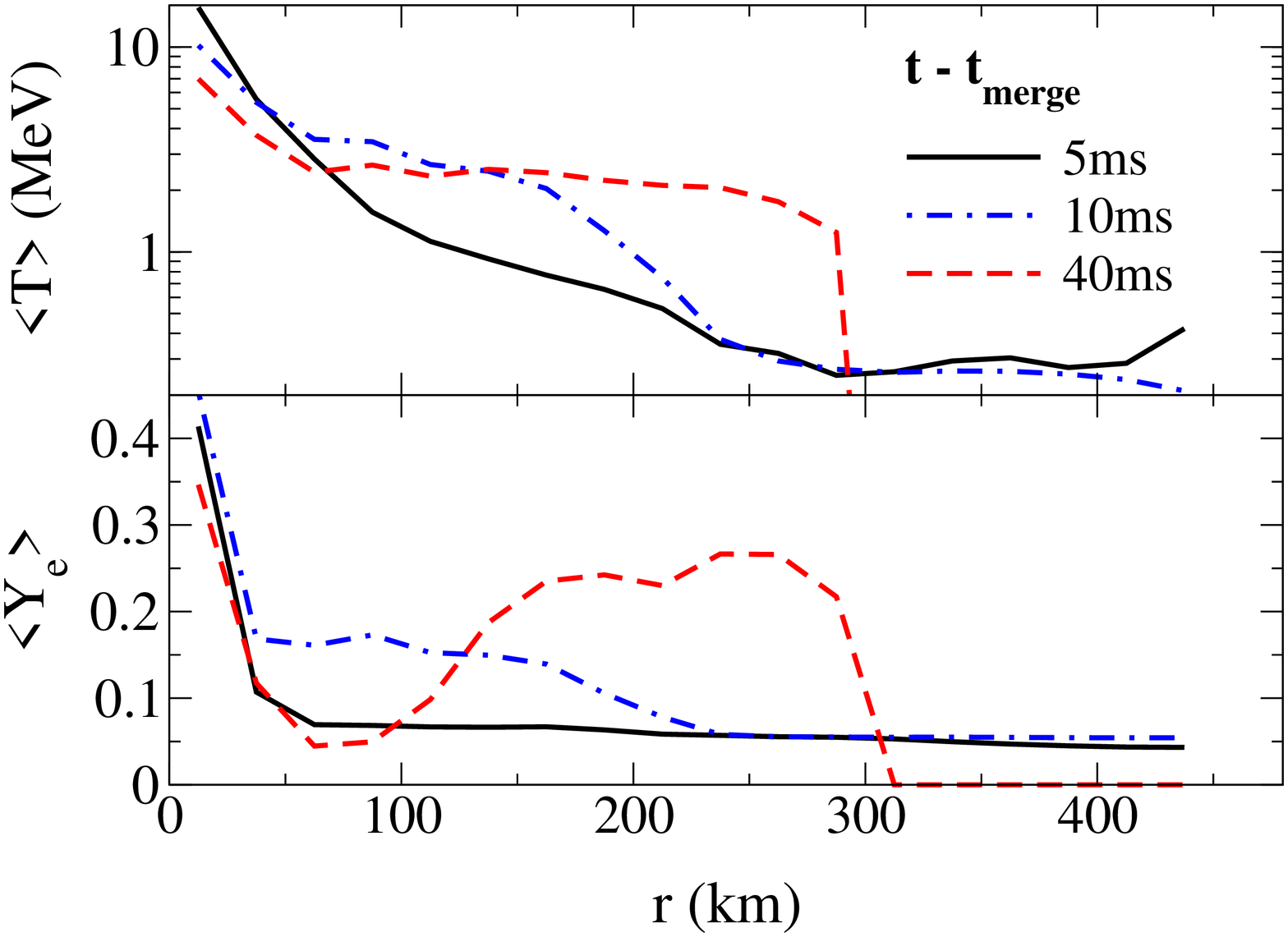}
\caption{Distribution of the average temperature and electron fraction around the remnant black hole
for simulation M14-7-S8, at the same times as in Fig.~\ref{fig:RhoVsR}. 
The average is taken after subdividing all material into radial
bins identical to those shown in Fig.~\ref{fig:RhoVsR}.}
\label{fig:RhoVsTYe}
\end{figure}

In the simulations with $M_{\rm BH}=7M_\odot$, we follow the post-merger evolution of the system.
We define the time of merger $t_{\rm merge}$ as the time at which $50\%$ of the neutron star has been accreted onto the black hole,
and the disk mass $M_{\rm disk}$ as the mass remaining in the accretion disk $20\,{\rm ms}$ after $t_{\rm merge}$.
We evolve the
disks for about $40\,{\rm ms}$ after $t_{\rm merge}$, switching to a fixed metric evolution after about $15\,{\rm ms}$.
We find a range of disk masses $M_{\rm disk}=0.04M_\odot\textrm{--}0.14M_\odot$. 
At late times, the accretion timescale in these disks is $t_{\rm acc}\sim 150\,{\rm ms}$ 
for the $M_{\rm NS}=1.2M_\odot$ simulations and $t_{\rm acc}\sim 75\,{\rm ms}$ for the $M_{\rm NS}=1.4M_\odot$ simulations.

\begin{figure*}
{\LARGE $t-t_{\rm merge}=5\,{\rm ms}$}\\
\includegraphics[width=0.55\columnwidth]{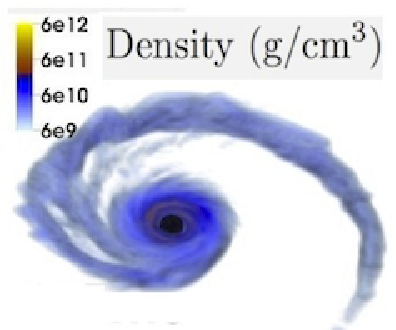}
\includegraphics[width=0.72\columnwidth]{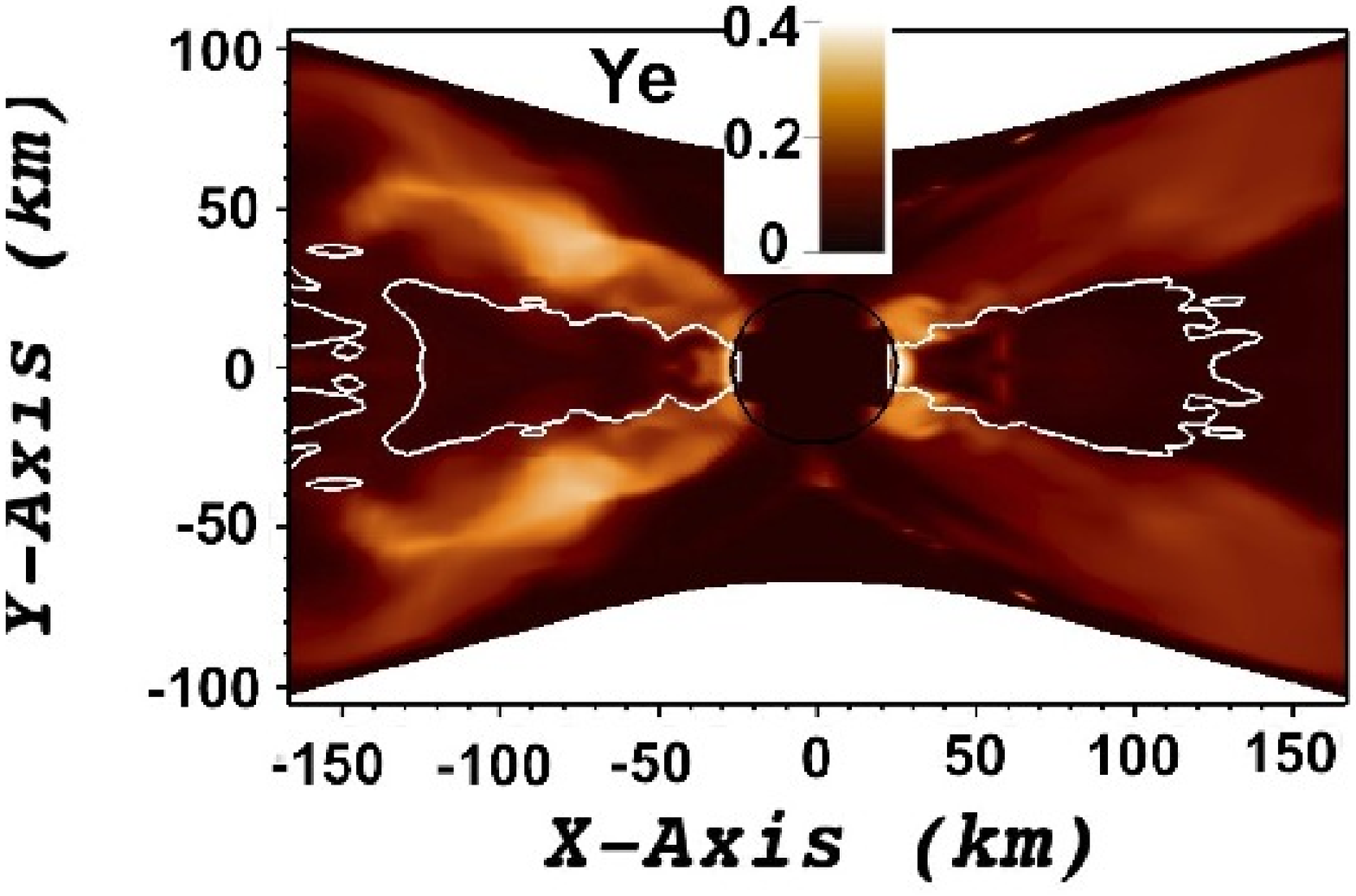}
\includegraphics[width=0.72\columnwidth]{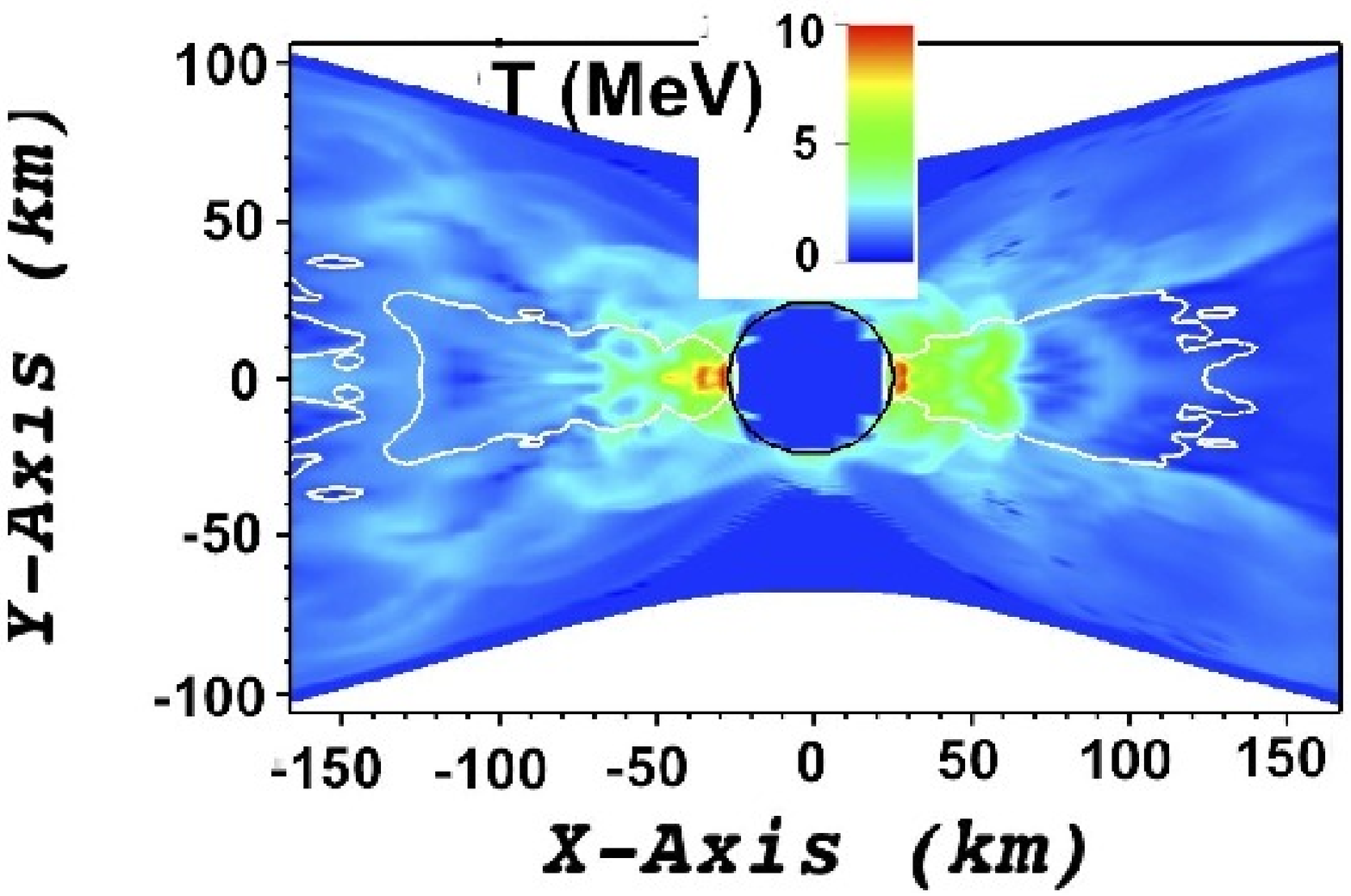}\\
\vspace{0.5cm}
{\LARGE $t-t_{\rm merge}=10\,{\rm ms}$}\\
\vspace{-0.2cm}
\includegraphics[width=0.55\columnwidth]{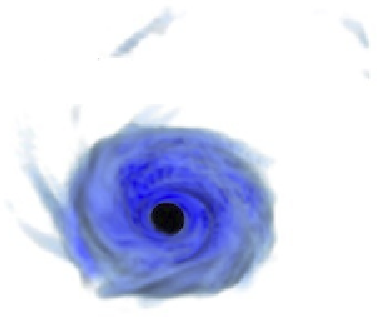}
\includegraphics[width=0.72\columnwidth]{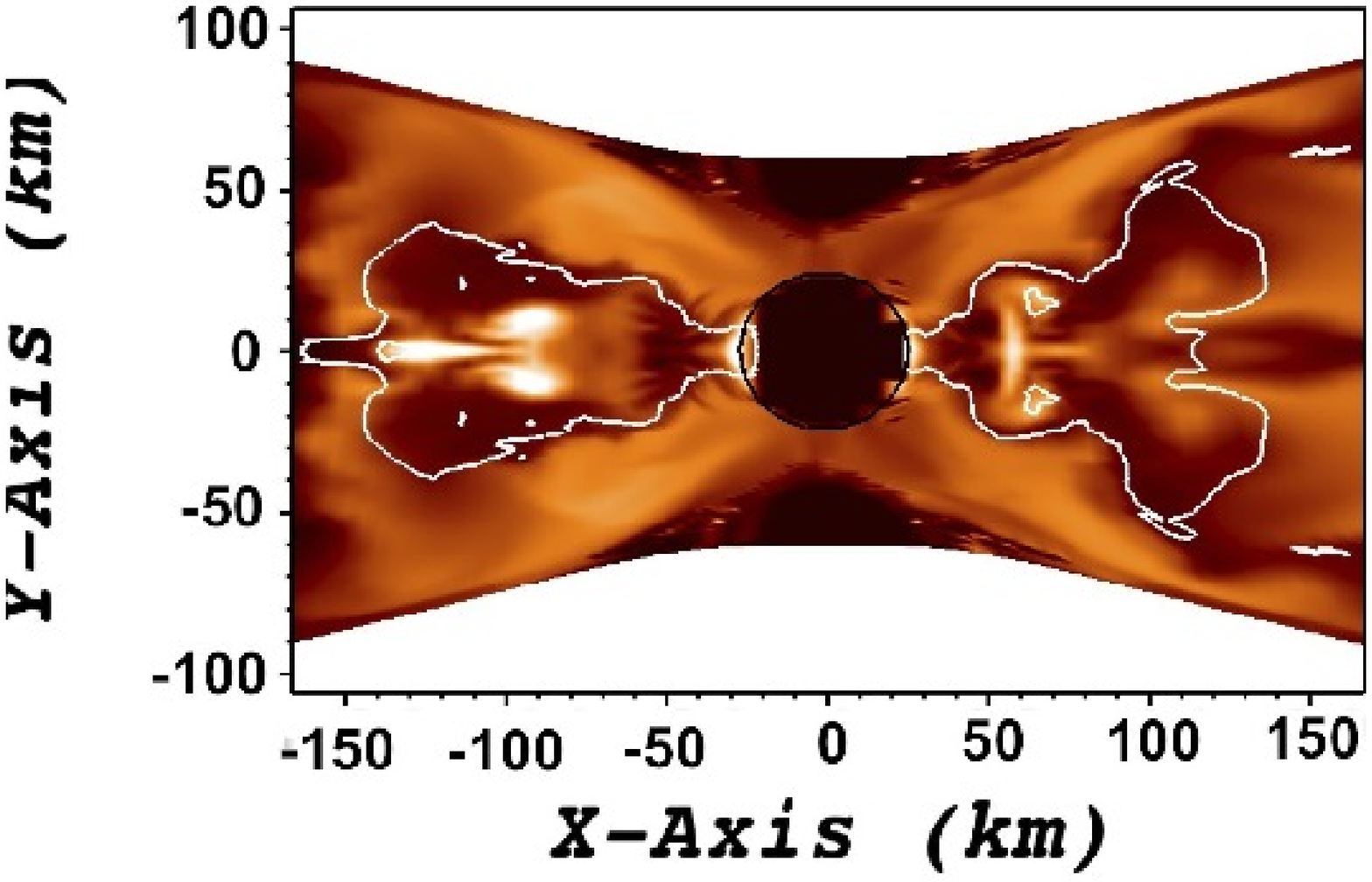}
\includegraphics[width=0.72\columnwidth]{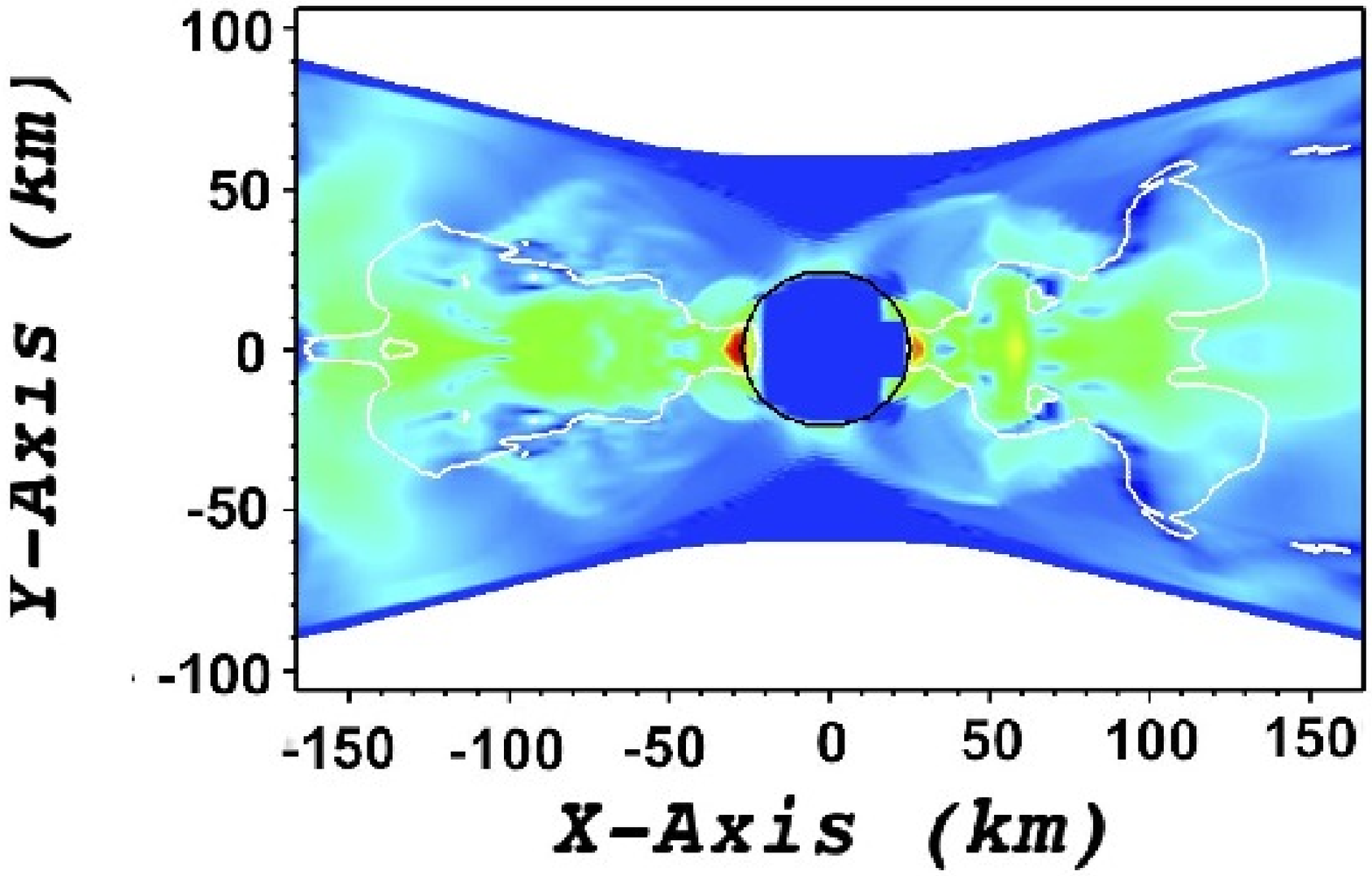}\\
\vspace{0.5cm}
{\LARGE $t-t_{\rm merge}=40\,{\rm ms}$}\\
\vspace{-0.2cm}
\includegraphics[width=0.55\columnwidth]{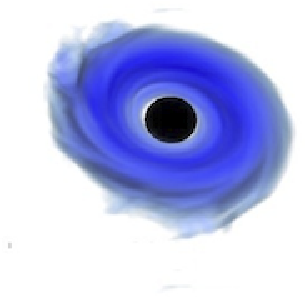}
\includegraphics[width=0.72\columnwidth]{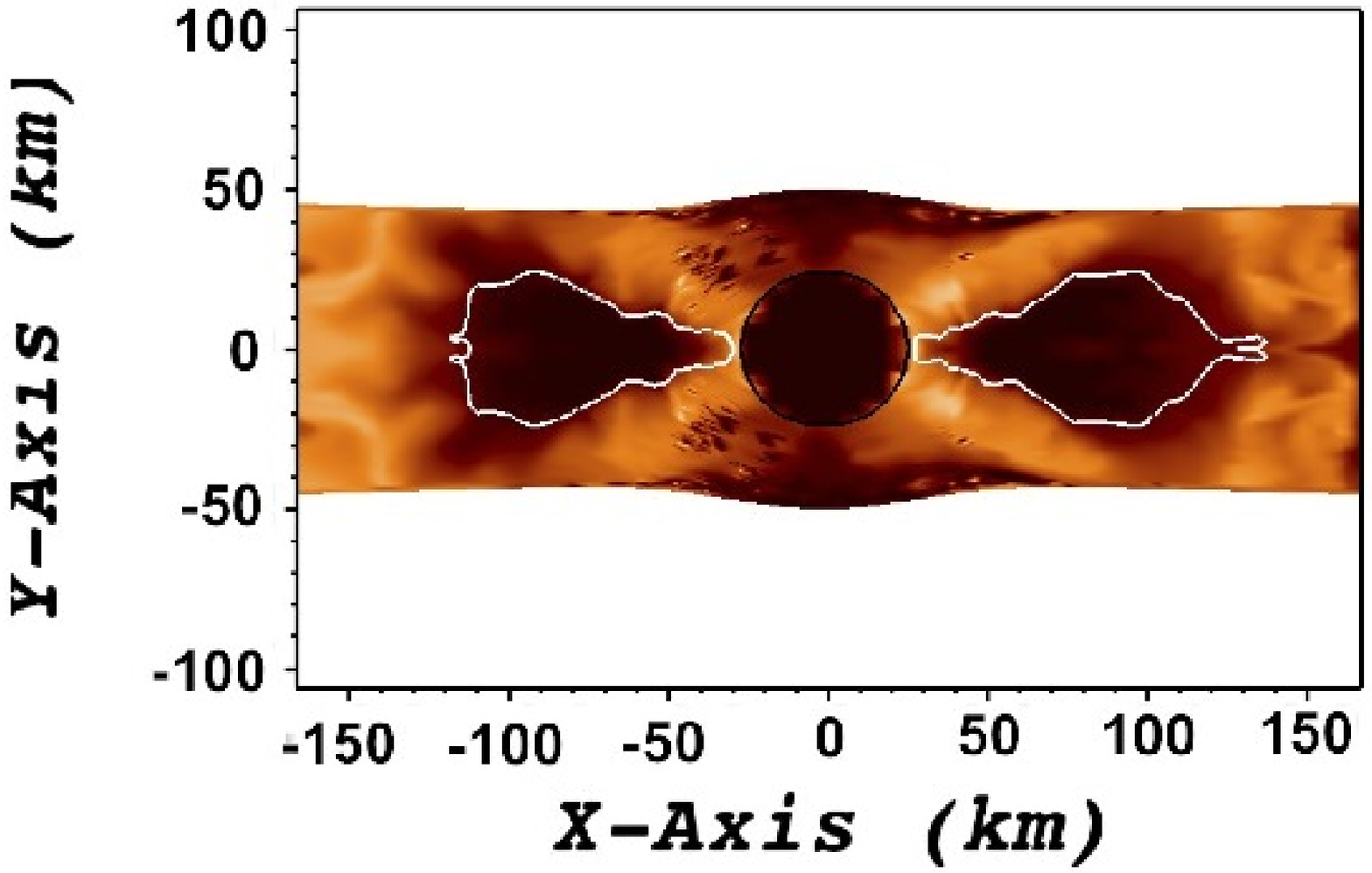}
\includegraphics[width=0.72\columnwidth]{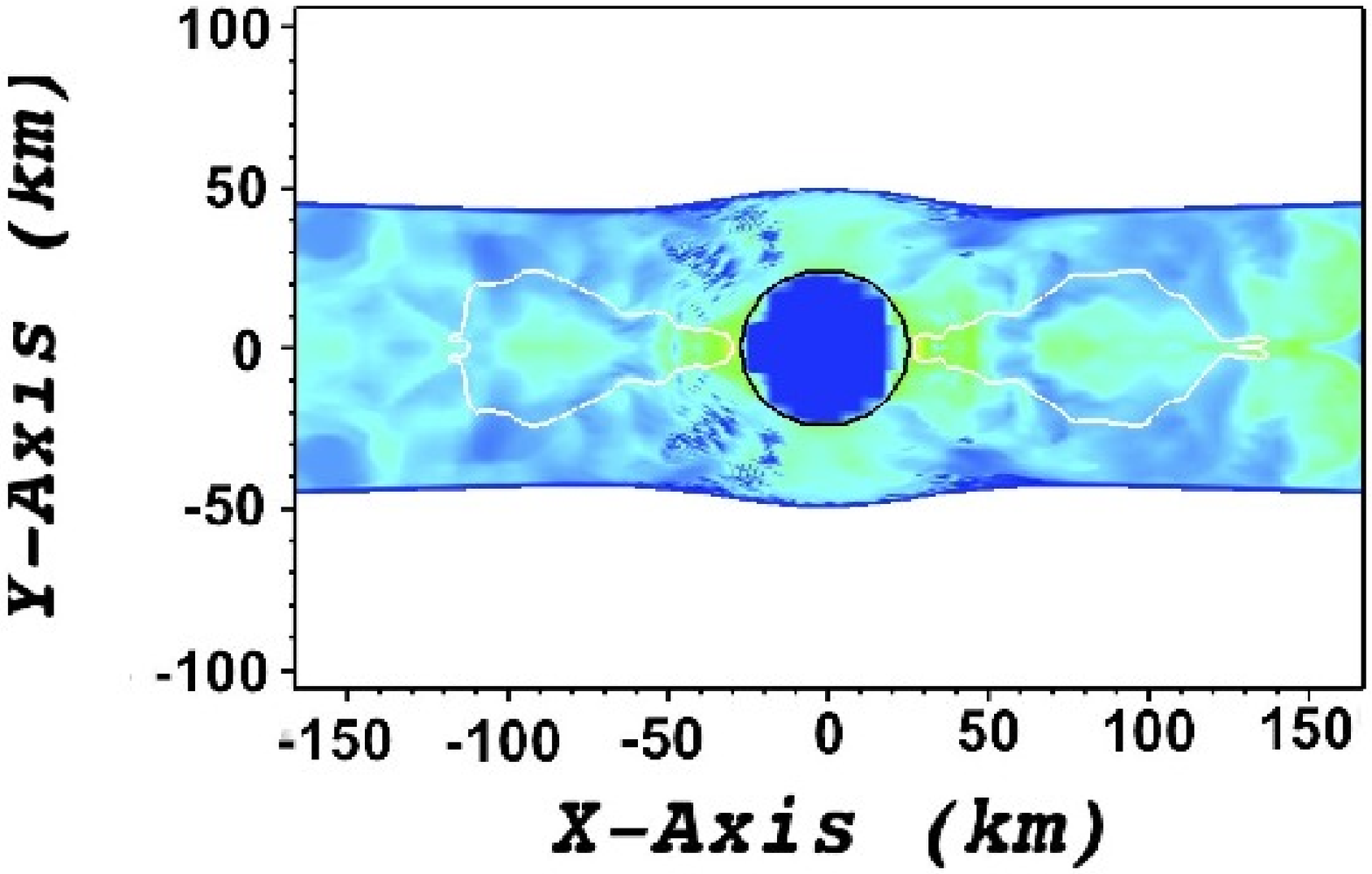}
\caption{Disk evolution for case M14-7-S8. The 3
snapshots are taken $5\,{\rm ms}$ after merger (top, unbound material
leaves the grid), $10\,{\rm ms}$ after merger (middle, end of rapid protonization of the disk), 
and $40\,{\rm ms}$
after merger (bottom, end of the simulation).
{\it Left:} 3D distribution of baryonic matter. {\it Center:}
Electron fraction in the vertical slice perpendicular to the orbital plane of the disk
which passes through the initial location of the black hole and neutron star. The white line is the
density contour $\rho=10^{10}{\rm g/cm^3}$. 
{\it Right: } Temperature in the same vertical slice. In the first two snapshots, the disk is hot and rapidly
protonizing. At later times, the disk cools down to $T\sim 2-3\,{\rm MeV}$ and becomes very neutron
rich again ($Y_e\sim 0.05$).}
\label{fig:DiskEvol}
\end{figure*}

The evolution of the disk can be subdivided into two main periods. For the first $5\,{\rm ms}\textrm{--}10\,{\rm ms}$, shocks in the forming
disk heat the material and raise the entropy of the fluid (see Fig.~\ref{fig:Entropy}). The resulting disk is compact and hot (see below).
However, average temperatures rapidly decrease
through a combination of effects: the expansion of the hot compact disk, the emission of neutrinos,
and the fall-back of cool material from the tail while hot material is accreted onto the black hole. An example
of this evolution is shown in Figs.~\ref{fig:RhoVsR} and \ref{fig:RhoVsTYe}, where we look in more detail at simulation
M14-7-S8. $5\,{\rm ms}$ after merger, about a third of the remaining material is in a compact hot disk, with most of the
mass at $r\sim 5M_{\rm BH}\sim 60\,{\rm km}$ and temperature $T\sim 5\,{\rm MeV}\textrm{--}15\,{\rm MeV}$. The rest is in a cool tidal tail, part of
which is unbound. $5\,{\rm ms}$ later, the disk has spread and slightly cooled down 
($T\sim 3\,{\rm MeV}\textrm{--}10\,{\rm MeV}$), while the fall-back rate from the tidal tail begins to decrease. 
At later times, and given that we do not include the effects of viscosity, we no longer have any significant source of 
heating in the disk. As hotter material accretes onto the black hole, cooler material falls onto the disk, and neutrinos
extract energy from the system, the disk cools down. At the end of the simulation, most of the disk
is at $T\sim 2\,{\rm MeV}$ (except for the small amount of hotter material at the inner edge of the disk). It is also confined to a
small radial extent of $50\,{\rm km}<r<100\,{\rm km}$. As discussed in Deaton et al. 2013~\cite{Deaton2013}, this is a direct
consequence of the loss of energy due to emitted neutrinos. Visualizations of the disk for the same simulation
are shown in Fig.~\ref{fig:DiskEvol}, at the same times. The evolution of the composition of the fluid will be discussed
in the next section, as it is tightly linked to the properties of the neutrino emission.

\begin{figure}
\flushleft
\includegraphics[width=0.85\columnwidth]{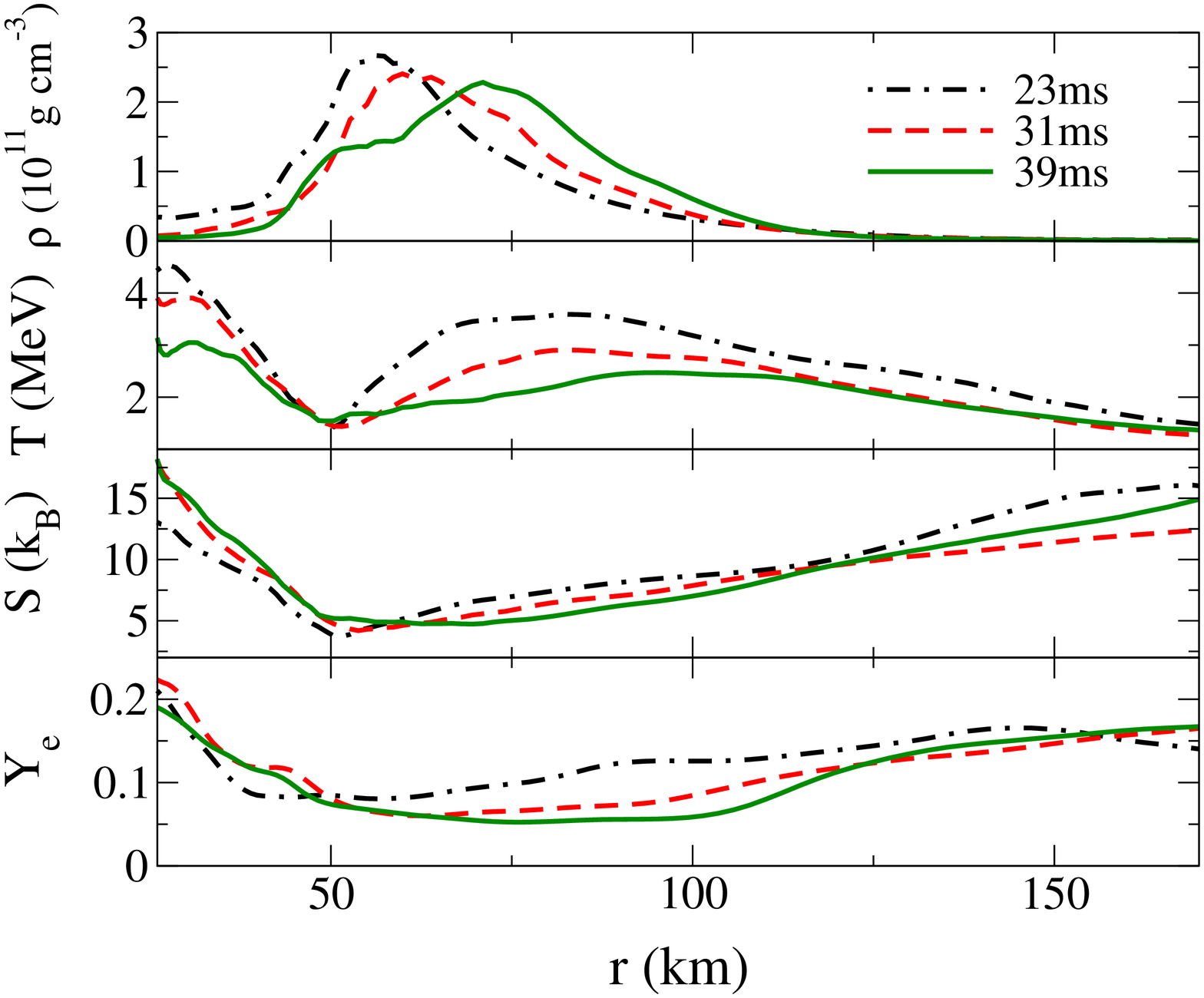}
\caption{The late-time post-merger evolution of the disk formed in case
M12-7-S8.  We plot density-weighted average values along cylindrical shells,
with the shell radius defined from the proper circumference of the cylinder
on the equator.  Data at three times are shown:  23, 31, and 39\,ms after
the merger time.}
\label{fig:DiskSettling}
\end{figure}

\begin{figure}
\flushleft
\includegraphics[width=0.85\columnwidth]{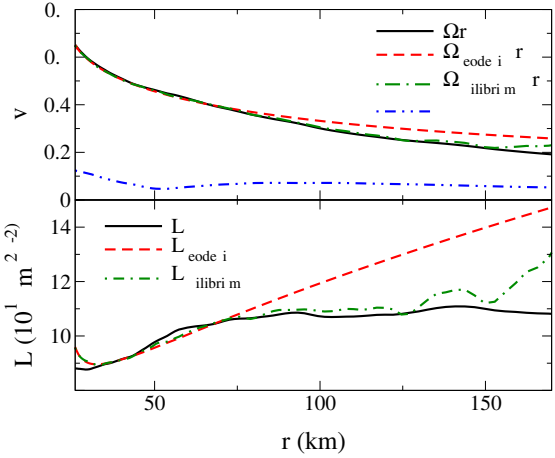}
\caption{The equatorial velocity in the $\phi$-direction, $v^{\phi}\equiv u^{\phi}r/u^t$, and the specific angular
  momentum $L\equiv -u_{\phi}/u_t$ in the disk for case M12-7-S8 measured
  31\,ms after merger. For comparison, the sound speed, $c_s$, is also included in the top panel,
  as are the $v^{\phi}$ and $L$ values for geodesic circular orbit and
  hydrodynamic equilibrium circular orbit for the given metric and pressure
  profile in both panels.}
\label{fig:DiskRotation}
\end{figure}

After the first 10\,ms, matter in the disks largely settles into circular
orbits, although they remain highly non-axisymmetric throughout the
post-merger evolution.  In Fig.~\ref{fig:DiskSettling}, we show
vertically- and azimuthally-averaged profiles of the density, temperature,
entropy, and electron fraction at three times, separated by 8\,ms, for
the case M12-7-S8. Other systems that produce large disks show very
similar profiles. The disk evolves under the influence of accretion
and neutrino emission, causing the disk to cool and the inner regions to
evacuate.  The $S$ and $Y_e$ profiles approach the ``U'' shape seen in
our previous simulation~\cite{Deaton2013}.  Fig.~\ref{fig:DiskRotation}
shows the angular velocity and angular momentum profiles 31\,ms after
merger.  The inner disk is almost entirely rotation-supported, as can
be seen from the agreement between the actual rotation and the curves
for a circular geodesic orbit.  Beyond 70\,km, the actual rotation is
significantly slower than geodesic, a sign that the outer disk has
significant pressure support.  This is confirmed by the agreement
between the actual rotation rate and what is required for a hydrodynamic
equilibrium circular orbit, an agreement that only breaks down in the
very outer regions where the disk has not equilibrated.  It is also
observed in the fact that the sound speed $c_s$ decreases much more slowly
with radius than the angular velocity.  The small change in $c_s$
is a consequence of  the moderate variation in temperature across the disk
and the very weak dependence of $c_s$ on density for the disk's range of
densities and temperatures.  The low
$\left.\frac{\partial c_s}{\partial\rho}\right|_{T,Y_e}$ comes from passing
through a minimum of $c_s$; at lower densities, the pressure is dominated
by relativistic particles, while at higher densities the degeneracy pressure
is stronger.  We observe that the angular
momentum of a freely falling particle on a circular orbit actually has
a minimum at
around 34\,km, which is also where the radial epicyclic frequency
of perturbed circular orbits (measured as in~\cite{2005Ap&SS.300..127A})
vanishes.  This is the innermost stable circular orbit (ISCO).  Inside the
ISCO, the fluid picks up a significant radially ingoing velocity
component, and, as would be expected, the angular momentum profile
is relatively flat.  It is worth pointing out the difference in this
inner-disk behavior from what was found in our previous
paper~\cite{Deaton2013}, which
involved a less-massive black hole and a more-massive accretion disk. 
In those simulations, we did not resolve a geodesic ISCO, but we did see
a sharp increase in the orbital energy in the inner disk, connected to
strong pressure forces.  This was associated with an apparent inner disk
instability.  No such effects are seen for these systems with more
massive black holes and less massive disks (see Sec.~\ref{sec:neutrino}).

\section{Neutrino cooling and disk composition}
\label{sec:neutrino}

Neutrino emission plays an important role in the post-merger evolution of NSBH remnants. It is indeed the main source of
energy loss for the resulting accretion disk, and may also deposit energy in the low-density regions above the disk through
$\nu\bar{\nu}$ annihilation, or drive outflows through neutrino absorption in the upper regions of the disk. 
The simple leakage scheme used in our simulations can only address the cooling of the disk, as we do not take into
account the effects of neutrino heating. A first guess at
the importance of $\nu \bar{\nu}$ annihilation and neutrino absorptions can however be made from the intensity of
the neutrino emission coming from the disk.

To understand the evolution of our disks under neutrino emission, it is helpful to compare our results to previous
simulations of accretion tori using a pseudo-Newtonian potential and with more controlled initial conditions
(Setiawan et al.\ 2006~\cite{Setiawan2006}). 
There are a few important differences between this paper's simulations and those
presented in~\cite{Setiawan2006}.  The approximate treatment of gravity used
in~\cite{Setiawan2006} might significantly affect the behavior of the inner
disk, especially for high-spin cases.  Also, the initial
conditions in~\cite{Setiawan2006} were axisymmetric (taking the azimuthal
average of the final result of an earlier NSBH merger simulation
performed with a Newtonian code~\cite{Janka1999}), were colder than
the initial post-merger state of our simulations
($T\sim 1\,{\rm MeV}\textrm{--}2\,{\rm MeV}$ in most of the disk), and used lower
black hole spin parameters (as the post-Newtonian potential performs
worse with higher spins).
Some of their simulations
also include an explicit viscosity, which in the most extreme cases ($\alpha=0.1$) causes the disk to heat to temperatures
similar to our initial conditions ($T\sim 10\,{\rm MeV}$). 
The typical density ($\rho_0\sim10^{10}\textrm{--}10^{11}{\rm g/cm^3}$) and electron fraction ($Y_e<0.05$) are fairly similar to those found
in our disks resulting from general relativistic mergers.

\begin{figure}
\flushleft
\includegraphics[width=0.85\columnwidth]{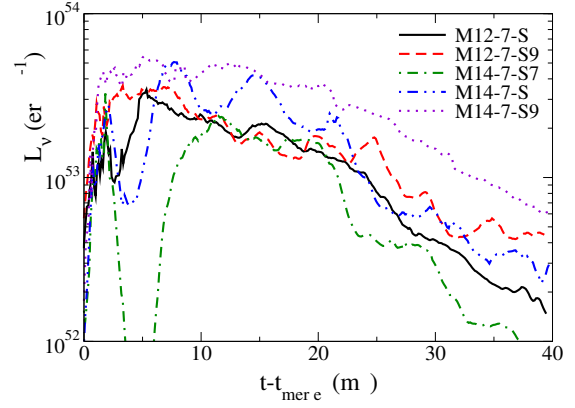}
\caption{Total neutrino luminosity (all species, at infinity) as a function of time,
for all simulations with $M_{\rm BH}=7M_\odot$.}
\label{fig:NuLum}
\end{figure}

\begin{figure}
\flushleft
\includegraphics[width=0.85\columnwidth]{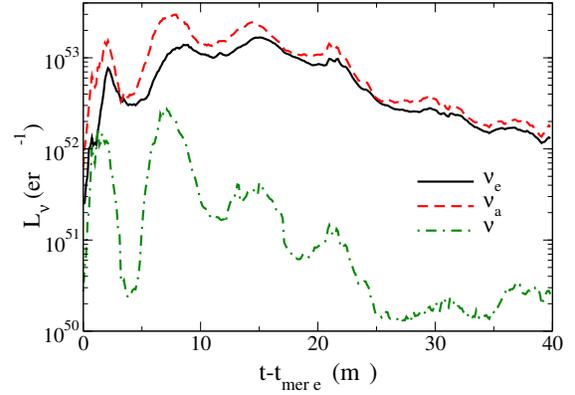}
\caption{Luminosity of the electron neutrinos, electron antineutrinos, and for each of the other
4 species ($\nu_{\mu,\tau}$, $\tilde\nu_{\mu,\tau}$), for simulation M14-7-S8.}
\label{fig:NuLumSpec}
\end{figure}

\begin{table*}
\caption{
Neutrino luminosities $L_\nu$ and average energy $\langle \epsilon_\nu \rangle$ $10\,{\rm ms}$ after merger. 
All luminosities are in unites of $10^{52}{\rm erg/s}$, and energies are in MeV. 
$L_{\nu_x}$ is the luminosity for each of the 4 types of neutrinos $\nu_{\mu,\tau},\bar{\nu}_{\mu,\tau}$ individually.
}
\label{tab:radiation}
\begin{tabular}{|c||c|c|c|c|c|c|}
\hline
Name & $L_{\nu_e}^{10{\rm ms}}$ & $L_{\bar{\nu}_e}^{10{\rm ms}}$ & $L_{\nu_x}^{10{\rm ms}}$ & $\langle \epsilon\rangle _{\nu_e}^{10{\rm ms}}$ & $\langle \epsilon \rangle_{\bar{\nu}_e}^{10{\rm ms}}$ & $\langle \epsilon \rangle_{\nu_x}^{10{\rm ms}}$ \\
\hline
M12-7-S8 & 9 & 13 & 0.2 & 11 & 14 & 16 \\ 
M12-7-S9 & 10 & 14 & 0.2 & 10 & 14 & 14 \\
M14-7-S7 & 6 & 11 & 0.2 & 12 & 16 & 18 \\
M14-7-S8 & 10 & 14 & 0.2 & 11 & 15 & 16 \\
M14-7-S9 & 13 & 23 & 1.5 & 11 & 14 & 16 \\
\hline
\end{tabular}
\end{table*}

\begin{figure}
\flushleft
\includegraphics[width=0.85\columnwidth]{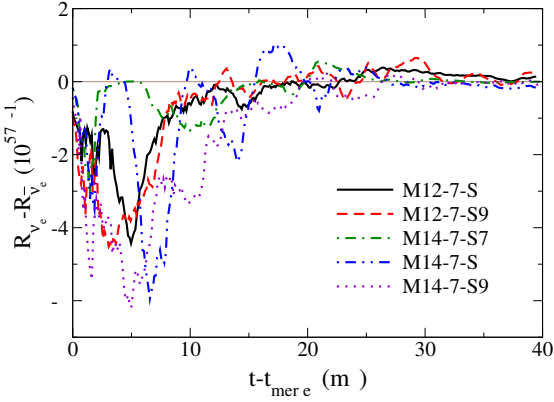}
\caption{Neutrino number emission as a function of time,
for all simulations with $M_{\rm BH}=7M_\odot$. 
A short period of rapid protonization occurs during
the first $10\,{\rm ms}\textrm{--}15\,{\rm ms}$ after merger. Then, the emission of electron neutrinos and antineutrinos become
more balanced.}
\label{fig:NuR}
\end{figure}

\begin{figure}\
\flushleft
\includegraphics[width=0.85\columnwidth]{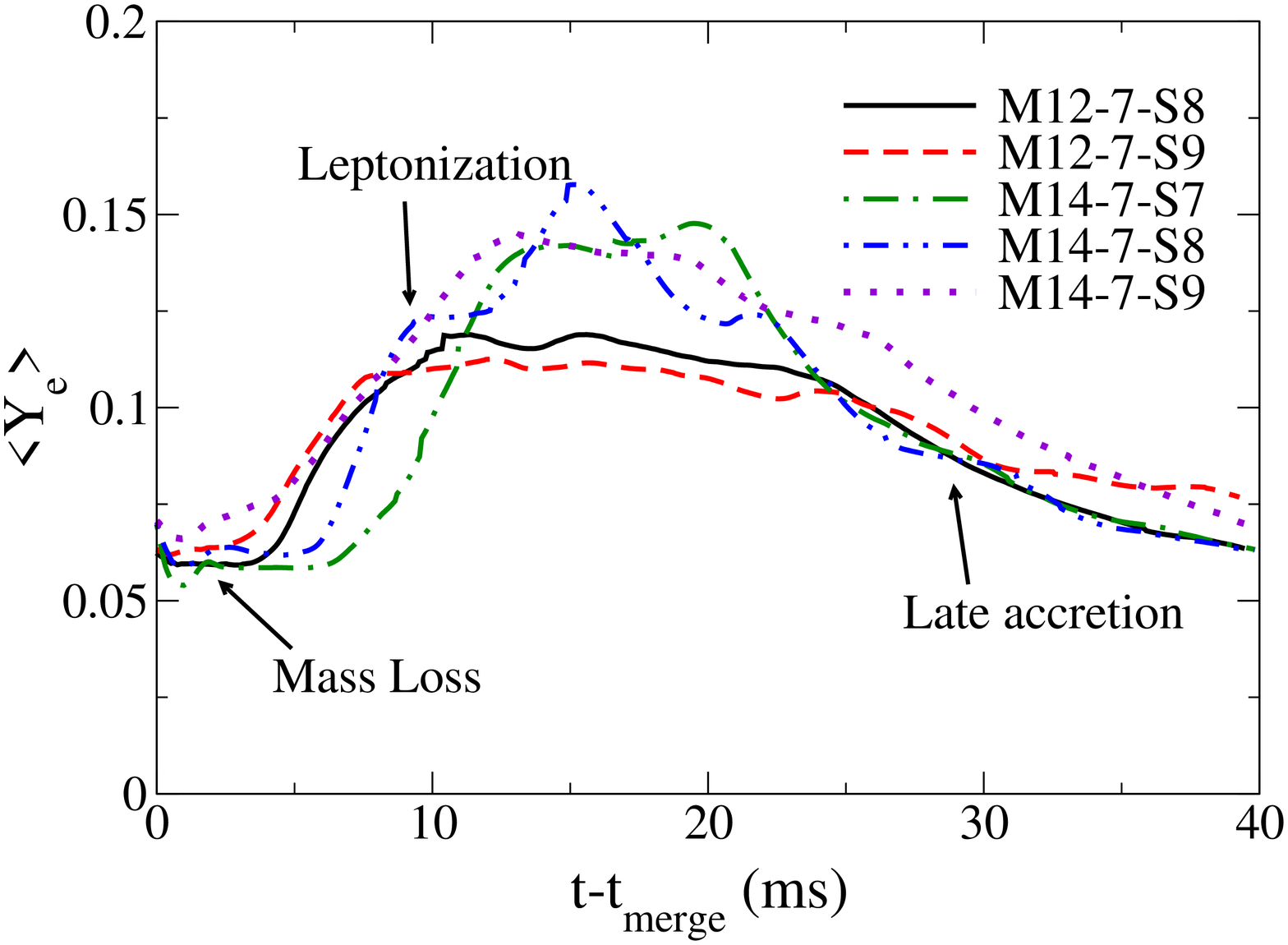}
\caption{Average electron fraction of the material on the grid,
for all simulations with $M_{\rm BH}=7M_\odot$. 
The early rise in $\langle Y_e\rangle$ is due to preferential emission
of electron antineutrinos, while the late time evolution is due to the accretion of the higher $Y_e$ material onto
the black hole and, for $M_{\rm NS}=1.2M_\odot$, the emission of a slight excess of electron neutrinos.}
\label{fig:Ye}
\end{figure}

\begin{figure}
\flushleft
\includegraphics[width=0.85\columnwidth]{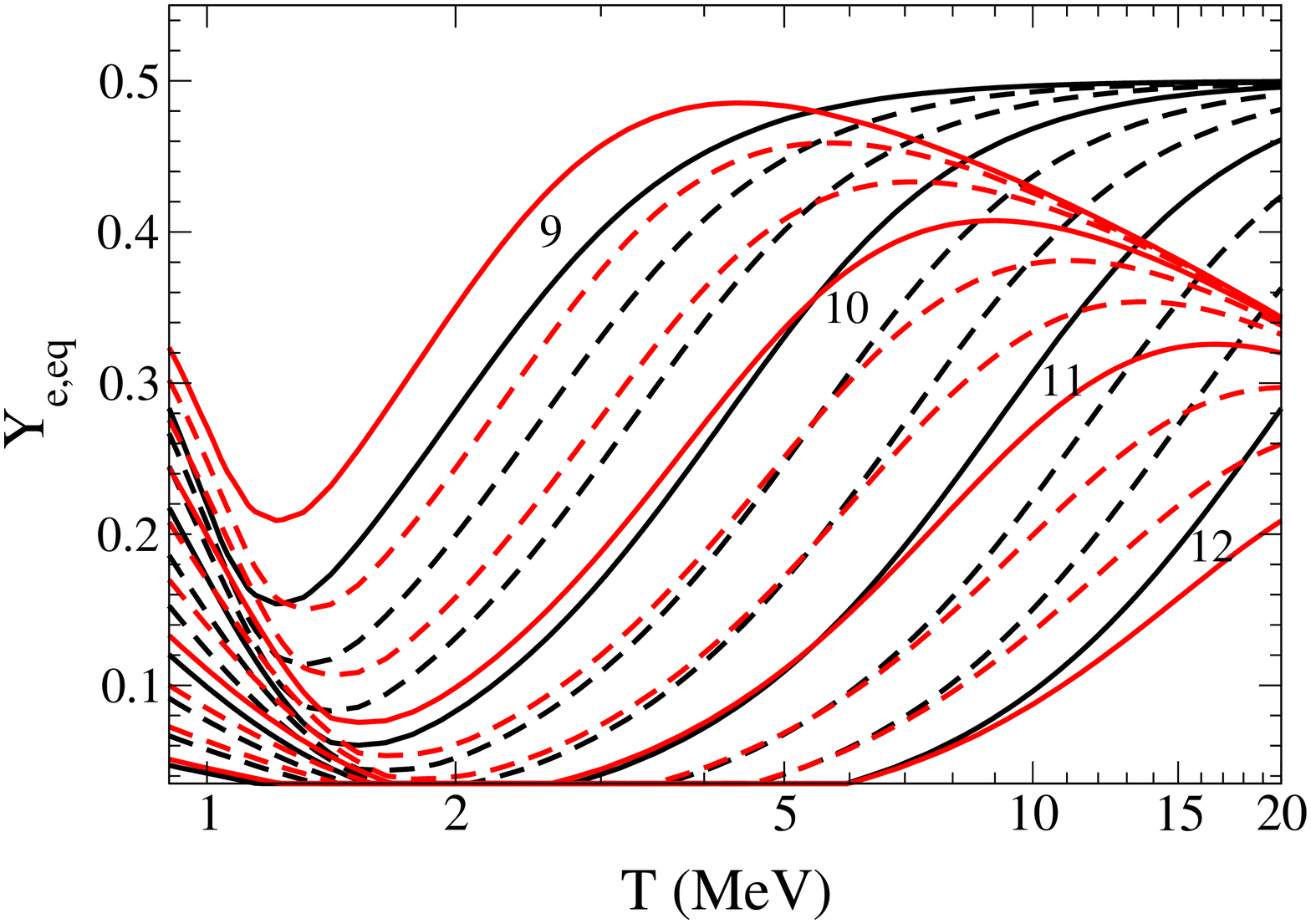}

\caption{Equilibrium electron fraction $Y_{e{\rm , eq}}$ in the free
  streaming regime, defined as the value of $Y_e$ at which the rate of
  positron and electron captures are equal for optically thin
  material, i.e.\ ignoring the effect of neutrino absorption on the
  matter. The black (thick) set of curves result from the balancing of
  the electron and positron capture rates used in our leakage code.
  There are corrections to these capture rates that make the energy
  dependence not $\propto E^2$. These cannot be included in our
  leakage scheme, but would adjust the value of $Y_{e{\rm , eq}}$ to
  the red set of curves. These corrections include those due to the
  neutron-proton rest mass difference, weak magnetism, and finite electron
  mass. The largest difference between these rates occurs either at
  low densities, where neutrino emission is low, or high temperatures,
  where the free streaming assumption is not valid. The various lines
  of each color show $Y_{e\rm , eq}(T)$ for a range of baryon
  densities (solid lines, labeled by $\log{\rho}=9,10,11,12$ in
  ${\rm g/cm^3}$). The dotted lines are separated by $\Delta
  \log{\rho}=1/3$.  We only show the range $0.035<Y_e<0.55$ covered by
  our equation of state table.}

\label{fig:YeEq}
\end{figure}

%\begin{figure}
%\flushleft
%\includegraphics[width=0.85\columnwidth]{NuEnergy}
%\caption{Average energy of the emitted neutrinos for electron neutrinos, electron antineutrinos, 
%and all other species for simulation M14-7-S8.}
%\label{fig:NuE}
%\end{figure}

For disks with $T\sim 10\,{\rm MeV}$, Setiawan et al.~\cite{Setiawan2006}
find total neutrino luminosities $L_\nu\sim 10^{53}{\rm erg/s}$, dominated by the emission of electron antineutrinos. This is expected
because the disk is both extremely neutron-rich and at a low enough density that, at this temperature, beta-equilibrium 
would require a more balanced distribution of neutrons and protons.
As a consequence, the disk rapidly protonizes, with the electron fraction rising to $Y_e\sim 0.1\textrm{--}0.3$ on a timescale of $20\,{\rm ms}$. 
This is fairly similar to the conditions observed in our disk at early times ($\sim 10\,{\rm ms}$ after merger), except that our initial conditions 
are asymmetric and that the high temperatures come from shock heating during disk formation rather than viscosity.
Accordingly, the early evolution of the disk and its neutrino emission follow a similar pattern. In Fig.~\ref{fig:NuLum}, we show the total neutrino
luminosity for all simulations with $M_{\rm BH}=7M_\odot$. At early times, we find $L_{\nu}\sim (2\textrm{--}5)\times 10^{53}{\rm erg/s}$, 
which appears consistent
with Setiawan et al.~\cite{Setiawan2006}, given the slightly higher temperatures present in our disks. As in~\cite{Setiawan2006}, the luminosity
also increases with the disk mass -- although the high variability of our disks makes it difficult to derive an actual scaling (it was $L_\nu \propto M_{\rm disk}^2$ in
~\cite{Setiawan2006}). The emission is dominated by electron antineutrinos, with $\tau$ and $\mu$ neutrinos having fairly negligible contributions to the total
luminosity (see Fig.~\ref{fig:NuLumSpec} for simulation M14-7-S8, and the summary of the properties of the neutrino emission for all simulations
in Table~\ref{tab:radiation}). The net lepton number emission, shown in Fig.~\ref{fig:NuR}, is highly variable but clearly causes a protonization of the disk -- an
effect confirmed by the evolution of the average electron fraction in the fluid (Fig.~\ref{fig:Ye}). Finally, the energy of the emitted neutrinos
is very consistent across all disks (see Table~\ref{tab:radiation}), with the electron neutrinos having the lowest energy
($\langle \epsilon_{\nu_e} \rangle \sim 10\,{\rm MeV}\textrm{--}12\,{\rm MeV}$),
and the electron antineutrinos being slightly more energetic ($\langle \epsilon_{\bar{\nu}_e} \rangle \sim 14\,{\rm MeV}\textrm{--}16\,{\rm MeV}$). 
The other species of neutrinos have energies
comparable to  $\langle \epsilon_{\bar{\nu}_e} \rangle$, are more variable, and are emitted at much lower rates. 
This is slightly lower than the neutrino energies observed for 
the hottest disks in~\cite{Setiawan2006} ($\langle \epsilon \rangle \sim 20\,{\rm MeV}$).
The large variations observed in the neutrino luminosity during the first $\sim 10\,{\rm ms}$ following the merger (see Fig.~\ref{fig:NuLum}) are due to the chaotic
nature of the disk formation process, and are particularly large for the lower mass disks. For example, simulation M14-7-S7 has a period $\sim 5\,{\rm ms}$ after
merger in which there is very little material in the hot inner disk (most of the material shocked at early times has been accreted by the black hole), and the neutrino
luminosity is very low. As the remaining material from the cold tidal tail falls onto the forming disk and heats up, the luminosity rises again. And once a
massive and more symmetric accretion disk forms, the variability in the neutrino luminosity decreases. Simulations with more massive post-merger remnants, on 
the other hand, rapidly form massive, hot inner disks and show less variability during these first $10\,{\rm ms}$.

%\begin{figure}
%\includegraphics[width=0.45\columnwidth]{LnueM14-7-S8-7105}
%\includegraphics[width=0.45\columnwidth]{LnuaM14-7-S8-7105}
%\caption{Loss of energy due to $\nu_e$ (left) and $\tilde\nu_e$ (right) 
%radiation as a function of volume,
%for simulation M14-7-S8 at the time of the 3rd snapshot in Fig.~\ref{fig:DiskEvol}}
%\label{fig:nuvol}
%\end{figure}

At later times, interesting changes in the properties of the disk arise due to the cooling of the disk and the fall-back of tail material. 
As hotter, more proton-rich material close to
the black hole is accreted, and cooler, more neutron-rich material ($Y_e<0.05$) from the tidal tail is added to the disk, we observe a decrease in the temperature
of the disk (see previous section), and in its average electron fraction.
Additionally, at the lower temperature observed in the disk after $20\,{\rm ms}\textrm{--}40\,{\rm ms}$ the matter is mainly optically thin.
In this regime, and under the assumption of no neutrino heating, the lepton number emission (and therefore the $Y_e$ evolution) is determined by the electron and positron capture rates alone.
The $Y_e$ where these rates are balanced, denoted as $Y_{e, \mathrm{eq}}$, as a function of density and temperature is shown in Fig.~\ref{fig:YeEq}.
For the conditions of our disks at late times ($T\sim2-3\,$MeV, $\rho\sim10^{11}\,$g/cm$^3$), the equilibrium value of $Y_e$ is lower than the instantaneous value of $Y_e \sim 0.1$.
Therefore, at that time, the neutrino emission from the optically thin parts of the disks is predominantly of positive lepton number and can lead to net deleptonization.
This is particularly visible in simulations M12-7-S8 and M12-7-S9, with a net excess of electron neutrinos over antineutrinos towards the end of the simulation (see Fig.~\ref{fig:NuR}).
In general, the neutronization of the disk is due to a combination of the accretion of neutron rich material and the preferential emission of electron neutrinos.
Note that, even at late times, the luminosity of the electron antineutrinos remains larger than the luminosity of the electron neutrinos, 
since their average energy is larger. 
Due to the steep dependence on temperature of the charged-current reactions $n + e^+ \rightarrow p + \bar{\nu}_e$ 
and $p + e^- \rightarrow n + \nu_e$, which
are the main contributors to the neutrino luminosity, the disk also becomes significantly dimmer as it cools down.

In nature, this evolution of the disk towards a lower electron fraction will eventually stop. Indeed, a physical disk will have a non-zero effective 
viscosity (presumably provided
by the magneto-rotational instability). When the heating due to viscous effects compensates the cooling due to neutrino emission, the disk should reach a more steady temperature
 profile. In the range of black hole masses
and spins studied here, we thus expect the neutrino emission to be composed of a short burst of mostly electron antineutrinos, with $L_\nu\sim 10^{53}{\rm erg/s}$, 
lasting $\sim 10\,{\rm ms}$ and followed by a period of more constant luminosity at a level set by the viscosity of the disk, decaying on a timescale comparable to the lifetime
of the disk. The emission should cause rapid protonization of the high-density regions of the disk in the first $10\,{\rm ms}$, but can then contribute to 
re-neutronization as the disk cools down.

For such a configuration, the evolution of $Y_e$ in the low-density regions above the disk (from which a wind might be launched) is expected, 
at early times, to be affected by the asymmetry in the number of electron neutrinos and antineutrinos emitted, and can slow down the protonization of the wind
material resulting from neutrino captures~\cite{2008ApJ...679L.117SR}.
The effect of the neutrino emission on a disk wind over longer timescales is more uncertain, 
particularly when general relativistic effects on the neutrino trajectories are taken into account. 
This is mainly because the antineutrinos can begin free-streaming much closer to the black hole than the neutrinos, for which the optical
depth is generally higher. Surman et al.~\cite{2013arXiv1312.1199S} have recently shown, based on neutrino fluxes derived from Newtonian simulations, that disk winds
could then be proton-rich, with an evolution resulting in the production of large amounts of ${\rm Ni}^{56}$ but no heavy elements ($A>120$), as opposed to the material unbound
during tidal disruption (see also~\cite{2011ApJ...743..155S}). Given the sensitivity of the process to the geometry of 
the neutrino radiation, which we do not directly compute, 
and the geometry of the wind, which we could only obtain by including neutrino heating, 
further studies are required to determine in which category the winds emitted by the disks produced in our simulations fall.

It is also worth noting some differences with the disk obtained for a lower mass black hole in Deaton et al. 2013~\cite{Deaton2013}.
In~\cite{Deaton2013}, the combination of a high black hole spin ($\chi_{\rm BH}=0.9$) and a low black hole mass ($M_{\rm BH}=5.6M_\odot$)
led to the formation of a massive disk ($M_{\rm disk}\sim 0.3M_\odot$). The disks presented here initially have optical depths
of a few ($\tau_{\nu_e}\sim 3\textrm{--}5 > \tau_{\bar{\nu}_e}$). Towards the end of the simulations, they have $\tau_{\nu_e}\sim 1$ and are
optically thin to electron antineutrinos. On the contrary, the more massive disk in~\cite{Deaton2013} remained optically
thick until the end of the simulation, $40\,{\rm ms}$ after merger (with $\tau_{\nu_e}\sim15$, $\tau_{\bar{\nu}_e}\sim 5$). 
The inner region of that disk was also susceptible to a convective instability, due to the negative gradient $dL/dr<0$ of the
angular momentum $L$ of the disk at radii $r\lesssim 27\,{\rm km}$, which prevented the disk to evolve towards an axisymmetric
configuration at late times. The less massive disks observed in this work, on the other hand, satisfy the Solberg-H{\o}iland criteria
for convective stability of an axisymmetric equilibrium fluid~\cite{1975ApJ...197..745S} everywhere outside of their ISCO.
More generally, the massive accretion
disk in Deaton et al.~\cite{Deaton2013} 
was hotter, denser, and had a higher electron fraction than any of the disks observed here. 
However, unless astrophysical black holes
are less massive than expected, or are very rapidly spinning ($\chi_{\rm BH} > 0.9$), 
we expect the less optically thick disks obtained in this paper to be more
representative of the post-merger remnants of NSBH binaries.

\section{Long-term evolution of the disk}
\label{sec:ltdisk}

Since our code does not include the effects of magnetic fields, or any ad-hoc prescription for the viscosity in the disk,
important physical effects will be missing if we attempt to evolve the post-merger disks over timescales comparable 
to the disk lifetime. From existing simulations
using approximate treatments of gravity~\cite{Setiawan2006,Lee:2009uc,Fernandez2013}, we can however obtain reasonable estimates 
of what should happen. As mentioned in the previous section, the cooling of the disk will stop when neutrino
cooling and viscous heating balance each other. For $\alpha\sim 0.01\textrm{--}0.1$, this is expected to lead to maximum
temperatures in the disk of $T_{\rm max}\sim 5\,{\rm MeV}\textrm{--}10\,{\rm MeV}$ and $Y_e\sim 0.1\textrm{--}0.3$~\cite{Setiawan2006}.
Over longer timescales, viscous transport of angular momentum will drive mass accretion onto the black hole, and
the viscous spreading of the disk. Over timescales of a few seconds, nuclear recombination and viscous heating
can then unbind $\sim 10\%$ of the matter in the disk~\cite{Fernandez2013} (see also~\cite{Lee:2009uc}). 
Although this ejecta is more
proton-rich than the material ejected during the disruption of the neutron star, 
Fernandez \& Metzger~\cite{Fernandez2013} estimate that its electron fraction ($Y_e\sim 0.2$) and 
specific entropy ($S\sim 8 k_B$) are low enough to allow the production
of 2nd and 3rd peak r-process elements. In NSBH mergers, this process would 
probably be of less importance than the prompt ejection of material: $10\%$ of the disk mass is much less
than the mass ejected at the time of merger, and likely difficult to detect if its properties are similar to
those of the dynamical ejecta. This is quite different from the results from BNS
mergers, in which only $10^{-4}M_\odot\textrm{--}10^{-2}M_\odot$ of material is promptly ejected, and the ejected material
is strongly heated by shocks and significantly affected by the neutrino radiation field~\cite{2014arXiv1402.7317W}: the different
components of the outflow can then be of similar importance, and have very different properties.

The exact mechanism leading to the production of relativistic jets from the post-merger remnants of NSBH
binaries is more uncertain. Mildly relativistic outflows along the spin-axis of the black hole have 
been obtained by Etienne et al.~\cite{Etienne:2012te}, but only after seeding a coherent poloidal field
$B_\theta\sim 10^{14}G$ in the disk resulting from a NSBH merger. For black hole spins initially
aligned with the orbital angular momentum of the binary (and thus aligned with the post-merger accretion disk),
the magnetic flux in the post-merger
remnant of a NSBH binary is likely to be too low to generate the highly efficient relativistic jets
which have been observed in numerical simulations of magnetically choked accretion flows~\cite{2012MNRAS.423.3083M},
at least until the accretion rate onto the black hole decreases by several orders of magnitude, or unless
the MRI can somehow lead to the growth of a coherent poloidal field.
But given the relatively high spins obtained after merger ($\chi_{\rm BH}\sim 0.9$), if such a flow can be established
it would have an efficiency $\eta=E_{\rm jet}/(\dot{M}c^2)\sim 100\%$~\cite{2012MNRAS.423.3083M}, or a jet
power $E_{\rm jet}\sim(10^{53}\,{\rm erg/s})\times(\dot{M})/(0.1\,M_\odot{\rm /s})$. Even if relativistic jets
can only be launched when the accretion rate drops to $\dot{M}\ll 1 M_\odot{\rm /s}$, this remains
sufficient to explain the energy output of short gamma-ray bursts. 
For black hole spins misaligned with the orbital angular momentum of the binary, the angular momentum 
of the post-merger accretion
disk can be misaligned by $5^{\circ}-15^{\circ}$ with respect to the final black hole spin~\cite{Foucart:2010eq,2013PhRvD..87h4053S}. 
Etienne et al.~\cite{Etienne:2012te} suggest that a large scale, coherent
poloidal field could be obtained from the mixing between poloidal and toroidal fields in a tilted accretion
disk, thus allowing the formation of a jet at earlier times. 
Unfortunately, the small number of existing general relativistic simulations of NSBH binaries with misaligned black hole 
spins~\cite{Foucart:2010eq,Foucart:2013a} do not include the effects of magnetic fields.

Finally, a few percents of the energy radiated in neutrinos is expected to be deposited in the region along
the spin axis of the black hole~\cite{Setiawan2006}, through $\nu \tilde{\nu}$ annihilations. This energy
deposition, which at early times in our disks would be $E_{\nu \tilde{\nu}}\sim 10^{51}\,{\rm erg/s}$,
might also be sufficient to power short gamma-ray bursts. 
Whether jets can be launched prompty or only after the accretion
rate drops, the disks produced by disrupting NSBH mergers for $M_{\rm BH}\sim 7M_\odot-10M_\odot$ thus appear
to be promising candidates for the production of short gamma-ray bursts.

\section{Gravitational Waves}
\label{sec:waves}

Given the relatively small number of orbits simulated in this work (5 to 8) and the lower phase accuracy
of simulations with a tabulated equation of state compared to simulations with a $\Gamma$-law equation of state, 
our simulations are not competitive with longer, more accurate recent results (e.g.~\cite{Foucart:2013a,Foucart:2013psa}) 
as far as the gravitational wave modeling of the inspiral phase is concerned. There are however two 
effects which can be studied even with these lower accuracy simulations: the influence of the tidal disruption of 
the neutron star on the gravitational wave signal, and the kick velocity given to the black hole as a result of an asymmetric
emission of gravitational waves around the time of merger.

\subsection{Neutron star disruption}
\label{sec:gwdisrupt}

The effect of the disruption of the neutron star on the gravitational wave signal generally falls into one out of three categories,
as shown in Kyutoku et al. 2011~\cite{Kyutoku:2011vz} (see also Pannarale et al. 2013~\cite{PannaraleEtAl2013} for a phenomenological
model of the waveform amplitude). When tidal disruption occurs far out of the innermost stable circular orbit (i.e.,
for low mass black holes), the spectrum of the gravitational wave signal shows a sharp cutoff at the disruption frequency.
At the opposite end, when the neutron star does not disrupt (that is, for higher mass black holes of low to moderate spins), 
the gravitational wave spectrum is similar to what is seen during the merger-ringdown of a binary black hole system (i.e.\ a bump
in the spectrum at the ringdown frequency, and an exponential cutoff at higher frequencies). Finally, for neutron stars
disrupting close to the innermost stable circular orbit or during their plunge into the black hole (for higher mass
black holes with large spin), the spectra show
a shallower cutoff of the signal at frequencies above the disruption frequency (in effect, a smaller contribution of the
ringdown of the black hole to the signal).

\begin{figure}
\flushleft
\includegraphics[width=0.85\columnwidth]{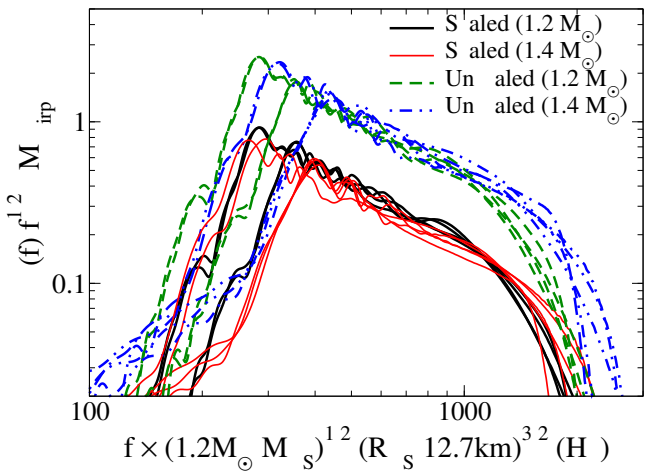}
\caption{Gravitational wave spectra for all simulations. Solid lines are rescaled in amplitude by the chirp mass 
$M_{\rm chirp}=(M_{\rm NS}M_{\rm BH})^{0.6}/(M_{\rm NS}+M_{\rm BH})^{0.2}$ and in frequency by the expected correction
to the tidal disruption frequency. Dashed curves show the spectra before rescaling in amplitude and
frequency. The amplitude is chosen arbitrarily (as it scales with the distance between the binary and the observer). 
Once rescaled, all spectra have roughly the same cut-off frequency.}
\label{fig:spectra}
\end{figure}

Most of the simulations presented here fall into the third category, as should be expected from the dynamics of the merger.
A simple approximation for the frequency of tidal disruption can be obtained by equating the tidal forces acting on the neutron
star with the gravitational forces on its surface, i.e. in Newtonian theory
\beq
\frac{M_{\rm BH}}{d_{\rm tide}^3} R_{\rm NS} \propto \frac{M_{\rm NS}}{R_{\rm NS}^2}\,\,,
\eeq
where $d_{\rm tide}$ is the binary separation at which tidal disruption occurs. We then obtain a gravitational wave frequency at
tidal disruption which scales as
\beq
\label{eq:f_gw_tide}
f_{\rm GW,tide} \propto \sqrt{\frac{G M_{\rm NS}}{R_{\rm NS}^3}}.
\eeq
We thus expect very similar high frequency signals for all simulations, with the cutoff occurring at a frequency $\sim 9\%$ larger
for simulations using a neutron star of mass $M_{\rm NS}=1.4M_\odot$ than for those using a neutron star of mass $M_{\rm NS}=1.2M_\odot$,
and the amplitude scaling with the chirp mass $M_{\rm chirp}=(M_{\rm NS}M_{\rm BH})^{0.6}/(M_{\rm NS}+M_{\rm BH})^{0.2}$.
Fig.~\ref{fig:spectra} shows that this is indeed the case -- the mass and spin of the black hole only contribute to the disruption
signal by determining how sharp the cutoff is, and not where the disruption starts. For the LS220 equation of state, we get
\beq
f_{\rm GW,tide} \sim 1.2\,{\rm kHz} \left(\frac{M_{\rm NS}}{1.2M_\odot}\right)^{1/2} \left(\frac{12.7\,{\rm km}}{R_{\rm NS}}\right)^{3/2}\,\,.
\eeq
The shape of the merger-ringdown waveform shows more variations for
the heavier neutron stars ($M_{\rm NS}=1.4M_\odot$), as simulations with $M_{\rm BH}=10M_\odot$ are close to the transition
between disrupting and non-disrupting systems.

\subsection{Kick velocity}
\label{sec:kick}

The second type of information that we can extract from the measured gravitational wave signal is the kick velocity given to the
black hole at merger as a result of asymmetric emission of gravitational waves. We find kicks spanning the 
range $v_{\rm kick}\sim 25\,{\rm km/s}\textrm{--}95\,{\rm km/s}$, with the expected approximate scaling of $v_{\rm kick}\propto q^{-2}$. 
These kick values are only slightly below what we would 
expect for binary black hole mergers with the same masses and spins (less than 20\% smaller, see~\cite{Lousto:2012gt} for the binary
black hole predictions). The kicks are however increasing with the black hole spin, while the opposite trend is observed in the fitting
formula for binary black holes when the spins are aligned with the orbital angular momentum and $q\gtrsim 4$.
In contrast, the difference with binary black hole results was shown to be much larger for lower mass black
holes~\cite{Foucart:2010eq}, which showed a stronger relative suppression of the gravitational recoil.

For disrupting NSBH binaries of large mass ratio and rapidly spinning black holes, however, 
those gravitational-wave induced kicks appear to be significantly weaker than the recoil velocities due to the asymmetric 
ejection of unbound material (see Sec.~\ref{sec:ejecta}).

\section{Conclusions}
\label{sec:conclusions}

We performed a first numerical study of NSBH mergers in the most likely range of black hole masses,
$M_{\rm BH}=7M_\odot\textrm{--}10M_\odot$,
using a general relativistic
code, a leakage scheme to estimate neutrino emission, and a hot nuclear-theory based equation of state (LS220).
We consider relatively high black hole spins $\chi_{\rm BH}=0.7\textrm{--}0.9$ 
and low mass neutron stars ($M_{\rm NS}=1.2M_\odot\textrm{--}1.4M_\odot$), 
so that the neutron star is disrupted by the
tidal field of the black hole before merger. Under those conditions, our simulations show that
NSBH mergers reliably eject large amounts of neutron-rich material, with 
$M_{\rm ej}=0.04M_\odot\textrm{--}0.20M_\odot$. The ejected material has a low entropy per baryon
$\langle S \rangle < 10k_B $, and is ejected early enough
that its composition is largely unaffected by neutrino emission and absorption. Accordingly, it should consistently produce
heavy r-process elements -- as opposed to BNS mergers, which show a wider variety of 
outflow compositions~\cite{2014arXiv1402.7317W}.

The mergers produce accretion disks of masses $M_{\rm disk}\sim 0.05M_\odot\textrm{--}0.15M_\odot$,
similar to the ejected masses.
Due to shock heating, these disks are initially hot ($T\sim 5\,{\rm MeV}\textrm{--}15\,{\rm MeV}$), and are
luminous in neutrinos ($L_{\nu}\sim 10^{53}{\rm erg/s}$). Preferential emission of electron 
antineutrinos causes the disks to rapidly protonize during the first $10\,{\rm ms}$ after merger,
from $Y_e\sim 0.06$ to $Y_e\sim 0.1\textrm{--}0.4$.
However, at later times, the cooling of the disk combined with the fall-back of cold, neutron rich
material from the tidal tail reverses this process. About $30\,{\rm ms}$ after merger, the denser
areas of the disks are already much colder ($T\sim 2\,{\rm MeV}\textrm{--}3\,{\rm MeV}$) and more neutron rich 
($Y_e<0.1$) than at early times. The disk is then expected to evolve mainly under the influence of MRI-driven
turbulence, and spread from its initial dense and compact state ($\rho\sim 10^{11}{\rm g/cm^3}$,
$r\sim 5M_{\rm BH}\sim 50\,{\rm km}$) over a viscous timescale of $t_{\rm \nu}\sim 0.1{\rm s}$. These
effects are however not modeled in our code, and we thus stop the simulations $40\,{\rm ms}$ after
merger. Outflows from these accretion disks are likely to be subdominant compared to the 
material dynamically ejected during the disruption of the neutron star. On the other hand, 
the disks are promising configurations for the production of short gamma-ray bursts.

Our simulations confirm that, for disrupting NSBH binaries, large recoil velocities 
$v_{\rm kick,ej} \gtrsim 100\,{\rm km/s}$ can be
imparted to the black hole. These recoil velocities are not due to the gravitational wave
emission, but rather to the asymmetry in the ejection of unbound material, as proposed
by Kyutoku et al.~\cite{Kyutoku:2011vz}. 

Finally, we showed that the disruption of a neutron star with a nuclear-theory based equation of
state has, for the larger black hole masses considered here, very different
properties from those observed for lower mass black holes or simpler equations of state.
The tidal tail is strongly compressed, to width of only a few kilometers, and the entire
disruption process occurs within a distance $r\lesssim 3M_{\rm BH}$ of the black hole (in the
coordinates of our simulation). In such cases, the properties of the post-merger remnant are
more sensitive to the initial parameters of the binary, and to the details of the equation
of state. Unfortunately, due to the differences of scale between the width of the tail, the radius of
the black hole, and the much larger distance to which we need to follow the ejecta, this also
makes these configurations significantly harder to resolve numerically. In the current version of 
our code, we can only evolve these mergers by using a very fine grid close to
the black hole, and letting the disrupted material exit the grid early. Improvements to the adaptivity
of our grid will be required in order for such binaries to be reliably evolved over
the same timescales as lower mass configurations.

This first study involved a fairly small number of cases, so many
questions have yet to be addressed.  Perhaps most interesting is
the variation in merger outcomes from different plausible assumptions
about the equation of state.  We have found that LS220 and $\Gamma=2$
results differ, but it would be much more interesting to
know if viably ''realistic'' equations of state can be distinguished
by their merger outcomes.  Also,
a more complete understanding of any of these NSBH post-merger evolutions
will require
simulations that include the remaining crucial physical effects. 
As ejecta continues to expand far from the merger location, its
density and temperature drop below the thresholds for maintaining
nuclear statistical equilibrium, and the subsequent compositional
and thermal evolution can only be modeled using nuclear reaction
networks. 
Additionally, tracking the disk evolution will require including the effects of MRI
turbulence, either directly
through MHD evolutions (the optimal solution), or at least indirectly
through some sort of effective viscosity prescription.
Finally, using radiation transport instead of a leakage scheme to evolve
the neutrino radiation field is needed to capture the production
of neutrino-driven winds, to estimate the deposition of energy and the production
of $e^+e^-$ pairs in the low-density region along the black hole
spin axis, and to generally improve the accuracy of our predictions
for the neutrino emission from the post-merger accretion disk.

\acknowledgments

The authors wish to thank Luke Roberts, Curran Muhlberger, Nick Stone, Carlos Palenzuela, Albino Perego,
Zach Etienne and Alexander Tchekhovskoy for useful discussions over the course of this
project, and the members of the SXS collaboration and the participants and organizers of the MICRA 2013 workshop
for their suggestions and support.
F.F.\ gratefully acknowledges support from the Vincent and Beatrice Tremaine Postdoctoral Fellowship. 
The authors at CITA gratefully acknowledge support from the NSERC Canada, from the Canada Research Chairs Program,
and from the Canadian Institute for Advanced Research.
M.D.D.\ and M.B.D.\ acknowledge support through NASA Grant No.\ NNX11AC37G and
NSF Grant PHY-1068243. 
L.K.\ gratefully acknowledges support from National Science Foundation (NSF) Grants No.\ PHY-1306125 and No.\ AST-1333129, while the authors at Caltech acknowledge support from NSF \
Grants No.\ PHY-1068881 and No.\ AST-1333520 and NSF CAREER Award No.\ PHY-1151197.
Authors at both Caltech and Cornell also thank the Sherman Fairchild Foundation for their support.
Computations were performed on the supercomputer Briar\'ee from the Universit\'e de Montr\'eal,
managed by Calcul Qu\'ebec and Compute Canada. The operation of this supercomputer is funded by the
Canada Foundation for Innovation (CFI), NanoQu\'ebec, RMGA and the Fonds de recherche du Qu\'ebec -
Nature et Technologie (FRQ-NT); and on the Zwicky cluster at Caltech,
supported by the Sherman Fairchild Foundation and by NSF award PHY-0960291. This work also used the Extreme Science and Engineering
Discovery Environment (XSEDE) through allocation No.\ TG-PHY990007N,
supported by NSF Grant No.\ ACI-1053575.

\bibliography{References/References}

\end{document}